\documentclass[12pt]{article}
\usepackage[utf8]{inputenc}
\usepackage{amsmath}
\usepackage{graphicx,psfrag,epsf}
\usepackage{enumerate}
\usepackage{bm}
\usepackage{caption}
\usepackage{color}
\usepackage[usenames,dvipsnames]{xcolor}
\usepackage{natbib}
\usepackage{subcaption}
\usepackage{url}
\usepackage{siunitx}
\usepackage{placeins}
\usepackage{rotating}
\usepackage{pdflscape}
\usepackage{verbatim}
\usepackage{listings}
\usepackage{hyperref}
\usepackage{multirow}
\usepackage{tikz}

\bibpunct{(}{)}{;}{a}{,}{,}

\newcommand{\blind}{0}

\newcommand{\E}{\mathop{\mbox{\sf E}}}
\def\defeq{\stackrel{\mathrm{def}}{=}}

\renewcommand{\thefootnote}{\fnsymbol{footnote}}

\hypersetup{
	colorlinks   = true,
	linkcolor    = black,
	citecolor    = black,
	urlcolor     = blue
}

\newenvironment{regtable5}[1]{
                \begin{table}[#1]
                               \sisetup{
                                               detect-all = false,
                                               table-number-alignment = center,
                                               add-integer-zero = false,
                                               table-figures-integer = 1,
                                               table-figures-decimal = 3,
                                               table-sign-mantissa = false,
                                               input-symbols = {()-},
                                               table-space-text-pre = -,
                                               table-space-text-post =***,
                                               bracket-negative-numbers = false,
                                               input-signs= +
                               }
                               \centering
	}{
\end{table}
}

\newenvironment{summarytable}[1]{
	\begin{table}[#1]
		\sisetup{
			detect-all,
			table-number-alignment = center,
			add-integer-zero = false,
			table-figures-integer = 1,
			table-figures-decimal = 3,
			round-mode = places,
			round-precision = 3
		}
		\centering
		
	}{
\end{table}
}

\begin{document}

\def\spacingset#1{\renewcommand{\baselinestretch}%
{#1}\small\normalsize} \spacingset{1}


\if0\blind
{
  \title{\bf Distillation of News Flow into Analysis of Stock Reactions*\let\thefootnote\relax\footnotetext{* This research was supported by the Deutsche Forschungsgemeinschaft through the SFB 649 `Economic Risk', Humbold-Universit\"{a}t zu Berlin. We like to thank the Research Data Center (RDC) for the data used in this study. We would also like to thank the International Research Training Group (IRTG) 1792.\\ This is a post-peer-review, pre-copyedit version of an article published in the Journal of Business and Economic Statistics. The final authenticated version is available online at: \url{http://dx.doi.org/10.1080/07350015.2015.1110525}}}
  \author{Junni L. Zhang\\
    Guanghua School of Management and Center for Statistical Science\\
    Peking University\\
    Beijing, 100871, China\\
     \\
    Wolfgang K. H\"{a}rdle \\
    Humboldt-Universit\"{a}t zu Berlin\\
    Unter den Linden 6, Berlin 10099, Germany\\
    and\\
    Sim Kee Boon Institute for Financial Economics\\
    Singapore Management University\\
    Administration Building, 81 Victoria Street, Singapore 188065\\
     \\
    Cathy Y. Chen\\
    Chung Hua University\\
    707, Sec.2, WuFu Rd., Hsinchu, Taiwan 30012\\
    and\\
    Humboldt-Universit\"{a}t zu Berlin\\
    Unter den Linden 6, Berlin 10099, Germany\\
    \\
    Elisabeth Bommes\\
    Humboldt-Universit\"{a}t zu Berlin\\
Unter den Linden 6, Berlin 10099, Germany}
  \maketitle
} \fi

\if1\blind
{
  \bigskip
  \bigskip
  \bigskip
  \begin{center}
    {\LARGE\bf Distillation of News Flow into Analysis of Stock Reactions}
\end{center}
  \medskip
} \fi

\bigskip
\newpage

\begin{abstract}
The gargantuan plethora of opinions, facts and tweets on financial business offers the opportunity to test and analyze the influence of such text sources on future directions of stocks.  It also creates though the necessity to distill via statistical technology the informative elements of this prodigious and indeed colossal data source.  Using mixed text sources from professional platforms, blog fora and stock message boards we distill via different lexica sentiment variables. These are employed for an analysis of stock reactions: volatility, volume and returns.
An increased sentiment, especially for those with negative prospection, will influence volatility as well as volume.  This influence is contingent on the lexical projection and different across Global Industry Classification Standard (GICS) sectors.  Based on review articles on 100 S\&P 500 constituents for the period of October 20, 2009 to October 13, 2014, we project into BL, MPQA, LM lexica and use the distilled sentiment variables to forecast individual stock indicators in a panel context.
Exploiting different lexical projections to test different stock reaction indicators we aim at answering the following research questions:\\
(i)	Are the lexica consistent in their analytic ability?\\
(ii) To which degree is there an asymmetric response given the sentiment scales (positive v.s. negative)?\\
(iii) Are the news of high attention firms diffusing faster and result in more timely and efficient stock reaction? \\
(iv) Is there a sector specific reaction from the distilled sentiment measures?\\
We find there is significant incremental information in the distilled news flow and the sentiment effect is characterized as an asymmetric, attention-specific and sector-specific response of stock reactions.
\end{abstract}

\noindent%
{\it Keywords: Investor Sentiment, Attention Analysis, Sector Analysis, Volatility Simulation, Trading Volume, Returns}  \\
\noindent%
{\it JEL Classifications:}  C81, G14, G17
\vfill

\newpage
\spacingset{1.45} 

\section{Introduction}
News are driving financial markets.  News are nowadays massively available on a variety of modern digital platforms with a wide spectrum of granularity scales.  It is exactly this combination of granularity and massiveness that makes it virtually impossible to process all the news relevant to certain financial assets.  How to distinguish between ``noise" and ``signal" is also here the relevant question.  With a few exceptions the majority of empirical studies on news impact work has therefore been concentrated on specific identifiable events like scheduled macroeconomic announcements, political decisions, or asset specific news.  Recent studies have looked at continuous news flow from an automated sentiment machine and it has been discovered to be relevant to high frequency return, volatility and trading volume.  Both approaches have limitations since they concentrate on identifiable indicators (events) or use specific automated linguistic algorithms.

This paper uses text data of different granularity from blog fora, news platforms and stock message boards.  Using several lexical projections, we define pessimistic (optimistic) sentiment with specific meaning as the average proportions of negative (positive) words in articles published in specific time windows before the focal trading day, and examine their impacts on stock trading volume, volatility and return.  We analyze those effects in a panel data context and study their influence on stock reactions. These reactions might be interesting since large institutions, more sophisticated investors, usually express their views on stock prospective or prediction through published analyst forecasts. However, analysts' recommendations may be contaminated by their career concerns and compensation scheme; they may also be in alliance with other financial institutions such as investment banks, brokerage houses or target companies \citep{Hong:2003, Liu:2012}. Due to the possible conflicts of interest from analysts and their powerful influence on naive small investors, the opinions from individual small investors may be trustworthy since their personal opinions hardly create any manipulation that governs stock reactions. The advent of social media such as {\it Seeking Alpha} enables small investors to share and express their opinions frequently, real time and responsively.
	
We show that small investors' opinions contribute to stock markets and create a ``news-driven" stock reaction. The conversation in the internet or social media is valuable since the introduction of conversation among a subset of market participants may have large effects on the stock price equilibrium \citep{Cao:2001}. Other literature such as \cite{Antweiler:2004},  \cite{Das:2007}, \cite{Chen:2014} demonstrate the value of individual opinions on financial market. They show that small investor opinions predict future stock returns and earnings surprises even after controlling the financial analyst recommendation.

The projections (of a text into sentiment variables) we employ are based on three sentiment lexica: the BL, LM and MPQA lexica. They are used to construct sentiment variables that feed into the stock reaction analysis. Exploiting different lexical projections, and using different stock reaction indicators we aim at answering the following research questions:\\
(i)	Are the lexica consistent in their analytic ability to produce stock reaction indicators, including volatility, detrended log trading volume and return?\\
(ii) To which degree is there an asymmetric response given the sentiment scales (positive v.s. negative)?\\
(iii) Are the news of high attention firms diffusing faster and result in more timely and efficient stock reaction? \\
(iv) Is there a sector specific reaction from the distilled sentiment measures?\\

Question (i) addresses the variation of news content across different granularity and lexica.  Whereas earlier literature focuses on numerisized input indices like ReutersNewsContent or Google Search Volume Index, we would like to investigate the usefulness of automated news inputs for e.g. statistical arbitrage algorithms.  Question (ii) examines the effect of different sentiment scales on stock reactions like volatility, trading volume and returns.  Three lexica are employed that are producing different numerical values and thus raise the concern of how much structure is captured in the resulting sentiment measure.  An answer to this question will give us insight into whether the well known asymmetric response (bad vs. good news) is appropriately reflected in the lexical projections.  Question (iii) and (iv) finally analyze whether stylized facts play a role in our study.  This is answered via a panel data scheme using GICS sector indicators and attention ratios.

\cite{Gros-Klusmann:2011} analyze in a high frequency context market reactions to the intraday stock specific ``Reuters NewsScope Sentiment" engine.  Their findings support the hypothesis of news influence on volatility and trading volume, but are in contrast to our study since they are based on a single news source and confined to a limited number of assets for which high frequency data are available.

\cite{Antweiler:2004} analyze text contributions from stock message boards and find that the amount and bullishness of messages have predictive value for trading volume and volatility. On message boards, the self-disclosed sentiment to hold a stock position is not bias free, as indicated in \cite{Zhang:2010}. \cite{Tetlock:2007} concludes that negative sentiment in a Wall Street Journal column has explanatory power for downward movement of the Dow Jones. \cite{Bollen:2011} classify messages from the micro-blogging platform Twitter in six different mood states and find that public mood helps to predict changes in daily Dow Jones values. \cite{Zhang:2012} extends this by filtering the Twitter messages (tweets) for keywords indicating a financial context and they consider different markets such as commodities and currencies.  \cite{Si:2013} use a refined filtering process to obtain stock specific tweets and conclude that topic based Twitter sentiment improves day-to-day stock forecast accuracy. \cite{Sprenger:2014}  also use tweets on stock level and conclude that the number of retweets and followers may be used to assess the quality of investment advice. \cite{Chen:2014} use articles and corresponding comments on {\it Seeking Alpha}, a social media platform for investment research, and show predictive value of negative sentiment for stock returns and earnings surprises. According to \cite{Wang:2014}, the correlation of Seeking Alpha sentiment and returns is higher than between returns and sentiment in Stocktwits, messages from a micro-blogging platform specialized in finance.

Using either individual lexical projections or a sentiment index comprising the common component of the three lexical projections, we find that the text sentiment shows an incremental influence on the stocks collected from S\&P 500 constituents. An asymmetric response of the stock reaction indicators to the negative and positive sentiments is confirmed and supports the leverage effect, that is, the stocks react to negative sentiment more. The reaction to the distilled sentiment measures is attention-specific and sector-specific as well. Due to the advent of social media, the opinions of small traders that have been ignored from past till now, do shed some light on stock market activity. The rest of the paper is organized as follows. Section \ref{sect:data} describes the data gathering process, summarizes definitions of variables and introduces the different sentiment lexica. In Section \ref{sect:EmpiricalResults}, we present the regression and simulation results using the entire sample and samples grouped by attention ratio and sectors. The conclusion follows in Section \ref{sect:Conclusion}.

\section{Data}
\label{sect:data}
\subsection{Text Sources and Stock Data}
While there are many possible sources of financial articles on the web, there are also legal and practical obstacles to clear before obtaining the data. The text source {\it Seeking Alpha}, as used in \cite{Chen:2014}, prohibits any application of automatic programs to download parts of the website (web scraper) in their Terms of Use (TOS). While the usage of web scrapers for non-commerical academic research is principally legal, these TOS are still binding as stated in \cite{Truyens:2014}.
For messages on Yahoo! Finance, another popular source of financial text data used in \cite{Antweiler:2004, Zhang:2010}, the TOS are not a hindrance but only limited message history is provided. As of December 2014, only the last 10,000 messages are shown in each stock specific message board and this roughly corresponds to a two-month-period for stocks that people talk frequently about like Apple.
In opposition to these two examples, NASDAQ offers a platform for financial articles by selected contributors including social media websites such as {\it Seeking Alpha} and {\it Motley Fool}, investment research firms such as Zacks. Neither do the TOS prohibit web scraping nor is the history of shown articles limited.
We have collected 116,691 articles and corresponding stock symbols, spanning roughly five years from October 20, 2009 to October 13, 2014. The data is downloaded by using a self-written web scraper to automate the downloading process. \\
\begin{figure}[ht!]
\begin{center}
\includegraphics[scale=0.8]{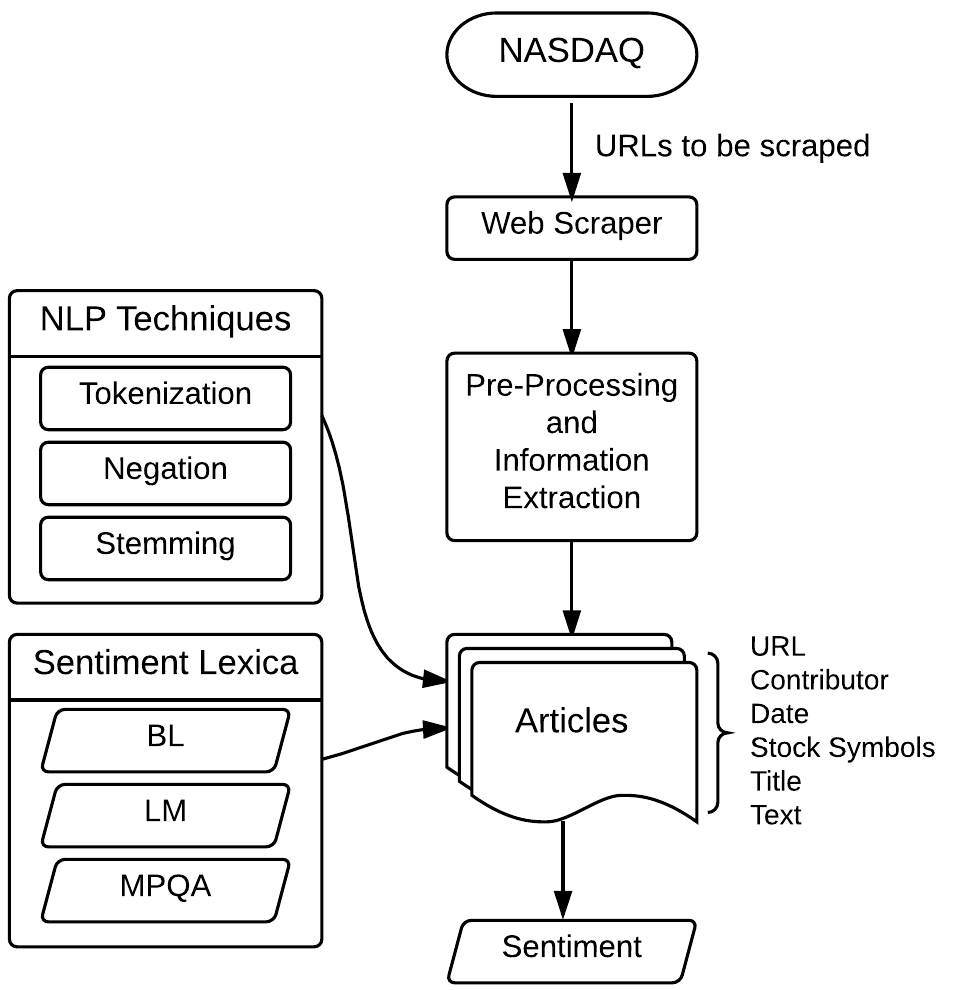}
\caption{Flowchart of data gathering process}
\label{fig:fc}
\end{center}
\end{figure}
The process of gathering and processing the article data and producing the sentiment scores can be seen in Figure \ref{fig:fc}.
Firstly, the URLs of all articles on NASDAQ are gathered and every webpage containing an article is downloaded. Each URL can be used in the next steps as unique identifier of individual articles to ensure that one article is not used twice due to real-time updates of the NASDAQ webpage. In the pre-processing step, the page navigation and design elements of NASDAQ are removed. The specifics of each article, namely contributor, publication date, mentioned stock symbols, title and article text, are identified and read out. In case of the article text, the results are stored in individual text files. This database is available for research purposes at \if0\blind{\href{https://sfb649.wiwi.hu-berlin.de/fedc/data.php}{RDC}, CRC 649, Humboldt-Universit\"{a}t zu Berlin}\fi \if1\blind{BLIND ITEM}\fi.

Furthermore, we collected stock specific financial data. Daily prices and trading volume, defined as number of shares traded, of all stock symbols that are S\&P 500 constituents are collected from Datastream while Compustat is used to gather the GICS sector for these stocks.

We consider three stock reaction indicators: log volatility, detrended log trading volume and return.  For stock symbol $i$ and trading day $t$, we first compute the \cite{Garman:1980} range-based measure of volatility defined as:
\begin{align}
\label{eq:Volatility}
 \sigma_{i,t}^2 &= 0.511 (u-d)^2 -0.019 \left \{c(u+d) - 2ud \right \}  - 0.383 c^2 \\
           \mbox{with} \ u &=  \log(P_{i,t}^H) - \log(P_{i,t}^O), \nonumber \\
                  d &= \log(P_{i,t}^L) - \log(P_{i,t}^O), \nonumber \\
                \ c &= \log(P_{i,t}^C) - \log(P_{i,t}^O), \nonumber
\end{align}
with $P_{i,t}^H$, $P_{i,t}^L$, $P_{i,t}^O$, $P_{i,t}^C$ as the daily highest, lowest, opening and closing stock prices, respectively.
\cite{Chen&etal:2006} and \cite{Shu&Zhang:2006} show that the Garman and Klass range-based measure of volatility essentially provides equivalent results to high-frequency realized volatility. For example, \cite{Shu&Zhang:2006} find that an empirical test with S\&P 500 index return data shows that the range-based variances are quite close to the high-frequency realized variance computed using the sum of 15-minute squared returns. \cite{Andersen:1997} show that the high-frequency realized volatility is very sensitive to the selected interval. In addition, it is also affected by the bid/ask spread. The range-based measure of volatility, on the other hand, avoids the problems caused by microstructure effects.
However, \cite{Alizadeh:2002} argue that range based measures such as the Garman-Klass estimator do not make use of the log-normality of volatility. As shown by \cite{Andersen:2001}, log realized volatility is less skewed and less leptokurtic in comparison to raw realized volatility.
Therefore, we use $\log \sigma_{i,t}$ instead, which also avoids regressing on a strictly positive variable in the subsequent analysis.

Following \cite{Girard:2007}, we estimate the detrended log trading volume for each stock by using a quadratic time trend equation:
\begin{equation}
\label{eq:Volume}
V^*_{i,t} = \alpha + \beta_1 (t-t_0) + \beta_2 (t-t_0)^2 + V_{i,t},
\end{equation}
 where $t_0$ is the starting point of the time window in consideration, $V^*_{i,t}$ is the raw daily log trading volume and the residual $V_{i,t}$ is the detrended log trading volume. In order to avoid imposing a look-ahead bias, for each trading day $t$, we use a rolling window of past 120 observations, $V^*_{i,t-120},\ldots,V^*_{i,t-1}$ with $t_0 = t - 120$, to estimate the coefficients and get a one-step ahead out-of-sample forecast $\hat{V}^*_{i,t}$, and then calculate $V_{i,t}=V^*_{i,t}-\hat{V}^*_{i,t}$.  Furthermore, we calculate the returns as $R_{i,t} = \log P_{i,t}^C - \log P_{i,t-1}^C$.

We focus on 100 stock symbols that are S\&P 500 constituents on all 1,255 trading days between October 20, 2009 and October 14, 2014, that belong to one of nine major GICS sectors for stock symbols that are S\&P 500 constituents on at least one trading day during this period, and that have the most trading days with articles. The distribution of GICS sectors among these 100 symbols are given in Table \ref{tab:sector}.  Out of the 116,691 articles collected, there are 43,459 articles associated with these 100 stock symbols; the number of articles for these stocks range from 340 to 5,435, and the number of trading days with articles ranges from 271 to 1,039. Most of the articles are not about one single symbol but contain references to several stocks.

\begin{table}[h]
\begin{center}
\begin{tabular}{lS}
\hline \hline
GICS Sector & {No. Stocks}\\ \hline
Consumer Discretionary & 21\\
Consumer Staples & 9\\
Energy & 6\\
Financials & 12\\
Health Care & 15\\
Industrials & 10\\
Information Technology & 21\\
Materials & 4\\
Telecommunication Services & 2\\ \hline \hline
\end{tabular}
\caption{Distribution of GICS sectors among the 100 stock symbols}
\label{tab:sector}
\end{center}
\end{table}

\subsection{Sentiment Lexica and Sentiment Variables}
To distill sentiment variables from each article, we use and compare three sentiment lexica.  The first lexicon (BL) is a list of 6,789 sentiment words (2,006 positive and 4,783 negative) compiled over many years starting from \cite{Hu&Liu:2004} and maintained by \href{http://www.cs.uic.edu/~liub/FBS/sentiment-analysis.html}{Bing Liu at University of Chicago, Illinois}.
We filter each article with this lexicon and calculate the proportions of positive and negative words.
The second lexicon (LM) is based on \cite{Loughran&McDonald:2011} which is specifically designed for financial applications, and contains 354 positive words, 2,329 negative words, 297 uncertainty words, 886 litigious words, 19 strong modal words and 26 weak modal words.  To be consistent with the usage of the other lexica, we only consider the list of positive and negative words and calculate the proportions of positive and negative words for each article.

The third lexicon is the MPQA (Multi-Perspective Question Answering) Subjectivity Lexicon by \cite{Wilson&etal:2005} which we later refer to as the MPQA lexicon.  This lexicon contains 8,222 entries.  In order to show the rather tedious distillation process let us look at six example entries:
\begin{lstlisting}[basicstyle=\ttfamily\scriptsize]
type=weaksubj  len=1  word1=abandoned  pos1=adj  stemmed1=n  priorpolarity=negative	
type=weaksubj  len=1  word1=abandonment  pos1=noun  stemmed1=n  priorpolarity=negative
type=weaksubj  len=1  word1=abandon  pos1=verb  stemmed1=y  priorpolarity=negative	
type=strongsubj  len=1  word1=abase  pos1=verb  stemmed1=y  priorpolarity=negative	
type=strongsubj  len=1  word1=abasement  pos1=anypos  stemmed1=y  priorpolarity=negative
type=strongsubj  len=1  word1=abash  pos1=verb  stemmed1=y  priorpolarity=negative	
\end{lstlisting}
Here \verb|type| refers to whether the word is classified as strongly subjective, indicating that the word is subjective in most contexts, or weakly subjective, indicating that the word only has certain subjective usages; \verb|len| denotes the length of the word; \verb|word1| is the spelling of the word; \verb|pos1| is part-of-speech tag of the word, which could take values adj (adjective), noun, verb, adverb, or anypos (any part-of-speech tag); \verb|stemmed1| is an indicator for whether this word is stemmed, where stemming refers to the process of reducing inflected (or sometimes derived) words to their word stem, base or root form; and \verb|priorpolarity| refers to polarity of the word, which could take values negative, positive, neutral, or both (both negative and positive).  The MPQA lexicon contains 4913 entries with negative polarity, 2718 entries with positive polarity, 570 entries with neutral polarity, and 21 entries with both polarity.  To be consistent with the usage of the other two lexica, we only consider positive and negative polarity.

We first use the NLTK package in Python to tokenize sentences and (un-stemmed) words in each article, and derive the part-of-speech tagging for each word.  We filter each tokenized article with the list of entries with \verb|stemmed1=n| in the MPQA lexicon to count the number of positive and negative word.  We then use the Porter Stemmer in the NLTK package to stem each word and filter each article with the list of entries with \verb|stemmed1=y| in the MPQA lexicon.  If a word has been assigned polarity in the first filtering step, it will no longer be counted in the second filtering step.  For each article, we can thus count the numbers of negative and positive words, and divide them by the length of the article to get the proportions of negative and positive words.

Regardless of which lexicon is used, we use a variation of the approach in \cite{Hu&Liu:2004} to account for sentiment negation.  If the word distance between a negation word (``not", ``never", ``no", ``neither", ``nor", ``none", ``n't") and the sentiment word is no larger than 5, the positive or negative polarity of the word is changed to be the opposite of its original polarity.

Among the words that appear at least three times in our list of articles, there are 470 positive and 918 negative words that are unique to the BL lexicon, 267 positive and 916 negative words that are unique to the LM lexicon, and 512 positive and 181 negative words that are unique to the MPQA lexicon.  The LM lexicon contains less unique positive words than the other two lexica, and the MPQA lexicon contains less unique negative words than the other two lexica.  Table \ref{tab:worduniq} presents the lists of ten most frequent positive words and ten most frequent negative words that are unique to these three lexica.  Since the BL and MPQA lexica are designed for general purpose and the LM lexicon is designed specifically for financial applications, the unique words under the BL and MPQA lexica indeed look more general.

Words in the general-purpose lexica may be misclassified for financial applications; for example, the word ``proprietary" in the negative list of the BL lexicon may refer to things like ``a secure proprietary operating system that no other competitor can breach" and hence have a positive tone in financial applications, and the word ``division" in the negative list of the MPQA lexicon may only refer to divisions of companies.  However, financial analysis using textual information is unavoidably noisy, and words in the LM lexicon can also be misclassified; for example, the word ``closing" in the negative list of the LM lexicon may actually refer to a positive event of closing a profitable deal.  Also, the LM lexicon does not take into account financial words such as ``debt" and ``risks" in the BL lexicon.

\FloatBarrier	
\begin{table}[ht]
	\centering
	\scalebox{0.75}{
	\centering
	\begin{tabular}{cc|cc|cc}
		\hline \hline
		\multicolumn{2}{c|}{BL} & 	\multicolumn{2}{c|}{LM} & 	\multicolumn{2}{c}{MPQA} \\
		{Positive (470)} & {Negative (918)} & {Positive (267)} & {Negative (916)} & {Positive (512)} & {Negative (181)}\\
		\hline
{Available} & {Debt} & {Opportunities} & {Declined} & {Just} & {Low} \\
(5,836) & (12,540) & (4,720) & (9,809) & (17,769) & (12,739) \\
{Led} & {Fell} & {Strength} & {Dropped} & {Help} & {Division} \\
(5,774) & (9,274) & (4,393) & (4,894) & (17,334) & (5,594) \\
{Lead} & {Fool} & {Profitability} & {Late} & {Profit} & {Least} \\
(4,711) & (5,473) & (4,174) & (4,565) & (15,253) & (5,568) \\
{Recovery} & {Issues} & {Highest} & {Claims} & {Even} & {Stake} \\
(4,357) & (3,945) & (3,409) & (3,785) & (13,780) & (4,445) \\
{Work} & {Risks} & {Greater} & {Closing} & {Deal} & {Slightly} \\
(3,808) & (2,850) & (3,321) & (3,604) & (13,032) & (3,628) \\
{Helped} & {Issue} & {Surpassed} & {Closed} & {Interest} & {Close} \\
(3,631) & (2,821) & (2,464) & (3,378) & (12,237) & (3,105) \\
{Enough} & {Falling} & {Enable} & {Challenges} & {Above} & {Trial} \\
(3,380) & (2,768) & (2,199) & (2,574) & (12,203) & (2,544) \\
{Pros} & {Aggressive} & {Strengthen} & {Force} & {Accord} & {Decrease} \\
(2,841) & (1,796) & (2,157) & (2,157) & (11,760) & (2,205) \\
{Integrated} & {Hedge} & {Alliance} & {Unemployment} & {Natural} & {Disease} \\
(2,652) & (1,640) & (1,842) & (2,062) & (10,135) & (2,001) \\
{Savings} & {Proprietary} & {Boosted} & {Question} & {Potential} & {Little} \\
(2,517) & (1,560) & (1,831) & (1,891) & (9,905) & (1,775) \\
		\hline \hline
	\end{tabular}
}
	\caption{Lists of ten most frequent positive words and ten most frequent negative words that are unique to the BL, MPQA or LM lexica, along with their frequencies given in parentheses.}
	\label{tab:worduniq}
\end{table}

We next investigate the pairwise relationship among the above three lexica.  Among the words that appear at least three times in our list of articles, there are 131 positive and 322 negative words that are shared only by the BL and LM lexica, 971 positive and 1,164 negative words that are shared only by the BL and MPQA lexica, and 32 positive and 30 negative words that are shared only by the LM and MPQA lexica.  It is not surprising that the two general-purpose lexica, BL and MPQA, share the most positive and negative words.  Out of the two general-purpose lexica, BL lexicon shares more positive and negative words with the special-purpose LM lexicon.  Table \ref{tab:wordshare} presents the lists of ten most frequent positive words and ten most frequent negative words that are shared only by two of these three lexica.  Words shared by the two general-purpose lexica (BL and MPQA) may be misclassified for financial applications; for example, the word ``gross" shared by the negative lists of these two lexica may refer to ``the annual gross domestic product" and have a neutral tone.  However, words shared by the LM lexicon and one of the general-purpose lexica may also be misclassified; for example, the word ``critical" shared by the negative lists of the BL and LM lexica may appear in sentences such as ``mobile devices are becoming critical tools in the worlds of advertising and market research" and have a positive tone.

\FloatBarrier	
\begin{table}[ht]
	\centering
	\scalebox{0.75}{
	\centering
	\begin{tabular}{cc|cc|cc}
		\hline \hline
		\multicolumn{2}{c|}{BL and LM} & 	\multicolumn{2}{c|}{BL and MPQA} & 	\multicolumn{2}{c}{LM and MPQA} \\
		{Positive (131)} & {Negative (322)} & {Positive (971)} & {Negative (1164)} & {Positive (32)} & {Negative (30)}\\
		\hline
{Gains} & {Losses} & {Free} & {Gross} & {Despite} & {Against} \\
(7,604) & (5,938) & (133,395) & (8,228) & (7,413) & (8,877) \\
{Gained} & {Missed} & {Well} & {Risk} & {Able} & {Cut} \\
(7,493) & (3,165) & (3,0270) & (7,471) & (5,246) & (3,401) \\
{Improved} & {Declining} & {Like} & {Limited} & {Opportunity} & {Challenge} \\
(7,407) & (3,053) & (24,617) & (5,884) & (4,398) & (1,042) \\
{Improve} & {Failed} & {Top} & {Motley} & {Profitable} & {Serious} \\
(5,726) & (2,421) & (14,899) & (5,165) & (3,580) & (1,022) \\
{Restructuring} & {Concerned} & {Guidance} & {Crude} & {Efficiency} & {Contrary} \\
(3,210) & (1,991) & (11,715) & (5,109) & (2,615) & (401) \\
{Gaining} & {Declines} & {Significant} & {Cloud} & {Popularity} & {Severely} \\
(3,150) & (1,654) & (10,576) & (4,906) & (1,588) & (348) \\
{Enhance} & {Suffered} & {Worth} & {Fall} & {Exclusive} & {Despite} \\
(2,753) & (1,435) & (10,503) & (4,732) & (1,225) & (342) \\
{Outperform} & {Weaker} & {Gold} & {Mar} & {Tremendous} & {Argument} \\
(2,518) & (1,288) & (9,303) & (3,190) & (611) & (324) \\
{Stronger} & {Critical} & {Support} & {Hard} & {Dream} & {Seriously} \\
(1,657) & (1,131) & (9,120) & (2,957) & (581) & (240) \\
{Win} & {Drag} & {Recommendation} & {Cancer} & {Satisfaction} & {Staggering} \\
(1,491) & (1,095) & (8,993) & (2,521) & (410) & (209) \\
		\hline \hline
	\end{tabular}
}
	\caption{Lists of ten most frequent positive words and ten most frequent negative words that are shared only by BL and LM lexica, only by BL and MPQA lexica, or only by LM and MPQA lexica, along with their frequencies given in parentheses.}
	\label{tab:wordshare}
\end{table}

The above discussion shows that projections using the three lexica are all noisy, therefore it is worthwhile to compare results from these projections.  For each stock symbol $i$ and each trading day $t$, we derive the sentiment variables listed in Table \ref{tab:variables} based on articles associated with symbol $i$ and published on or after trading day $t$ and before trading day $t+1$.
\begin{table}[h]
	\centering
	\scalebox{0.95}{
		\begin{tabular}{l|l}
			\hline \hline
			Sentiment Variable & Description\\\hline
			$I_{i,t}$ & \ Indicator for whether there is an article.\\
			$Pos_{i,t}$ (BL) & \ The average proportion of positive words using the BL lexicon.\\
			$Neg_{i,t}$ (BL) & \ The average proportion of negative words using the BL lexicon.\\
			$Pos_{i,t}$ (LM)& \ The average proportion of positive words using the LM lexicon.\\
			$Neg_{i,t}$ (LM)& \ The average proportion of negative words using the LM lexicon.\\
			$Pos_{i,t}$ (MPQA)& \ The average proportion of positive words using the MPQA lexicon.\\
			$Neg_{i,t}$ (MPQA)& \ The average proportion of negative words using the MPQA lexicon.\\\hline\hline
		\end{tabular}
		}
		\vspace{-6pt}
		\caption{Sentiment variables for articles published on or after trading day $t$ and before trading day $t+1$.}
		\label{tab:variables}
\end{table}

\section{Empirical Results}
\label{sect:EmpiricalResults}
\subsection{Entire Sample Results}
\subsubsection{Descriptive Statistics and Comparison of the Lexical Projections}
Table \ref{tab:SummaryTextVariables} presents summary statistics of the sentiment variables derived using the BL, LM and MPQA lexical projections for 43,569 symbol-day combinations with $I_{i,t}=1$, where $I_{i,t}$ is defined in Table \ref{tab:variables} and indicates whether there is an article associated with symbol $i$ and published on or after trading day $t$ and before trading day $t+1$.  This number is slightly different from the number of articles associated with the 100 selected symbols (43,459), since an article can be associated with multiple symbols.  The positive proportion is the largest under the MPQA projection, and the smallest under the LM projection.  The negative proportions under the three projections are similar.
Polarity in Table \ref{tab:SummaryTextVariables} measures the relative dominance between positive sentiment and negative sentiment. For example, the situation, $Pos_{i,t}$ (BL)$>Neg_{i,t}$ (BL), accounts for 88.04\% of the 43,569 observations. Note that under each projection, there are a small percentage of the observations for which $Pos_{i,t}=Neg_{i,t}$. Under both the BL and MPQA projections, positive sentiment is more dominant and widespread than negative sentiment.  The LM projection, however, results in a relative balance between positive and negative sentiment.
\FloatBarrier
\begin{summarytable}{ht}
	\centering
	\begin{tabular}{l|SSSSSSS[round-precision = 2, table-figures-decimal = 2]}
		\hline \hline
		Variable & {$\widehat{\mu}$} & {$\widehat{\sigma}$} & {Max} & {Q1} & {Q2} & {Q3} & {Polarity} \\
		\hline
		$Pos_{i,t}$ (BL)    & 0.033 & 0.012 & 0.134 & 0.025 & 0.032 & 0.040 & 88.04\% \\
		$Neg_{i,t}$ (BL)    & 0.015 & 0.010 & 0.091 & 0.008 & 0.014 & 0.020 & 10.51\% \\
		$Pos_{i,t}$ (LM)    & 0.014 & 0.007 & 0.074 & 0.009 & 0.013 & 0.018 & 55.70\% \\
		$Neg_{i,t}$ (LM)    & 0.012 & 0.009 & 0.085 & 0.006 & 0.011 & 0.016 & 40.17\% \\
		$Pos_{i,t}$ (MPQA)  & 0.038 & 0.012 & 0.134 & 0.031 & 0.038 & 0.045 & 96.26\% \\
		$Neg_{i,t}$ (MPQA)  & 0.013 & 0.008 & 0.133 & 0.007 & 0.012 & 0.017 &  2.87\% \\
		\hline
        \hline
		\multicolumn{8}{l}{ \rule{0pt}{2ex} \parbox{5.5in}{ \vspace{3pt} \footnotesize{Note: Sample mean, sample standard deviation, maximum value, 1st, 2nd and 3rd quartiles, and polarity. These descriptive statistics are conditional on $I_{i,t}=1$.}}}
	\end{tabular}
	\caption{Summary Statistics for Text Sentiment Variables}
	\label{tab:SummaryTextVariables}
\end{summarytable}

To check whether the sentiment polarity actually reflects the sentiment of the articles, we actually carefully checked and read the contents of 100 randomly selected articles and manually classified their polarity (positive, negative and neutral), and also use the lexical projections to automatically classify these articles as follows.  If the proportion of positive words for an article is larger than (or small than, or equal to) the proportion of negative words for the same article, then this article is automatically classified as positive (or negative, or neutral).
Table \ref{tab:sentclassification} reports the results.  It appears that the BL and MPQA projections put too much weight on positive sentiment, and are not powerful in detecting negative sentiment.  In contrast, the LM sentiment is powerful in detecting negative sentiment, but is not so good in detecting positive sentiment.

\begin{table}
		\centering
		\begin{tabular}{l|ccccccccc|c} \hline \hline
Manual& \multicolumn{3}{c}{BL Label} & \multicolumn{3}{c}{LM Label} & \multicolumn{3}{c}{MPQA Label} &\\
Label &  {Pos} & {Neg} & {Neu} & {Pos} & {Neg} & {Neu} & {Pos} & {Neg} & {Neu} & {Total} \\\hline
{Pos} & 56 & 4 & 1 & 41 & 12 & 8 & 61 & 0 & 0 & 61\\
{Neg} & 9 & 2 & 1 & 0 & 9 & 3 & 9 & 2 & 1 & 12\\
{Neu} & 22 & 5 & 0 & 10 & 15 & 2 & 26 & 0 & 1 & 27\\\hline
{Total} & 87 & 11 & 2 & 51 & 36 & 13 & 96 & 2 & 2 & 100\\
		\hline
        \hline
	\end{tabular}
	\caption{Sentiment Classification Results for 100 Randomly Selected Articles}
	\label{tab:sentclassification}
\end{table}

\begin{figure}[ht!]
	\begin{center}
		\includegraphics[scale=1]{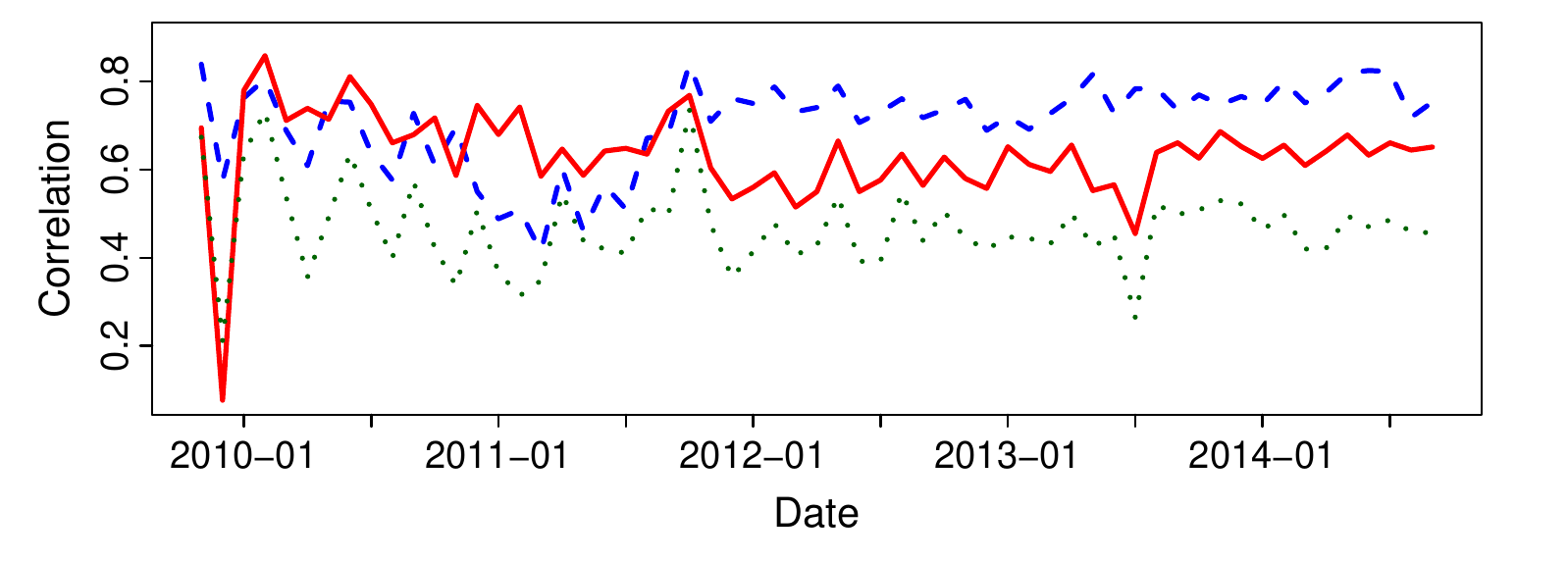}
		\caption{Monthly Correlation between Positive Sentiment: { \color{red}BL and LM (solid)}, { \color{blue}BL and MPQA (dashed)}, { \color{OliveGreen}LM and MPQA (dotted)}}
		\label{fig:sent_pos}
	\end{center}
\end{figure}

\begin{figure}[ht!]
	\begin{center}
		\includegraphics[scale=1]{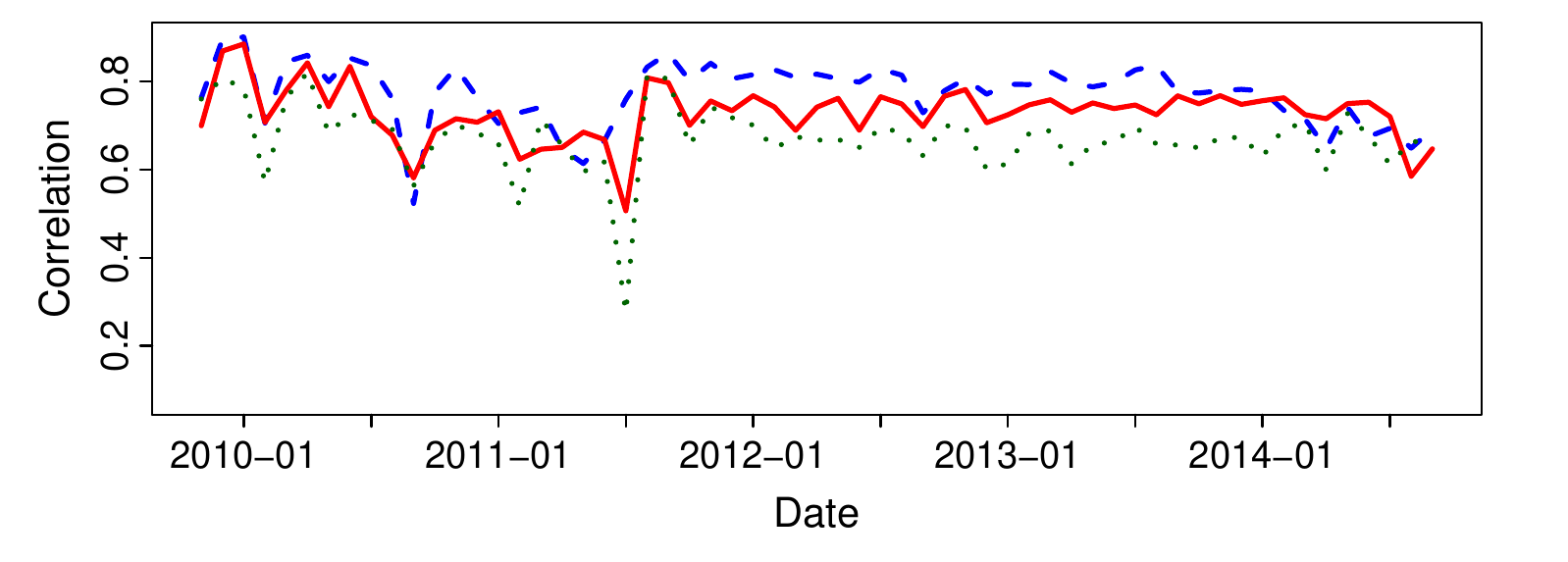}
		\caption{Monthly Correlation between Negative Sentiment: { \color{red}BL and LM (solid)}, { \color{blue}BL and MPQA (dashed)}, { \color{OliveGreen}LM and MPQA (dotted)}}
			\label{fig:sent_neg}
		\end{center}
	\end{figure}

Figure \ref{fig:sent_pos} and \ref{fig:sent_neg} respectively show the monthly correlation between positive and negative proportions under two of the three projections.  In general, the negative proportions are more correlated than positive proportions. Also, the correlation between the BL and LM projections and that between the BL and MPQA projections are larger than the correlation between the LM and MPQA projections, which is consistent with the discussion about the list of words shared by two of the three projections (see Table \ref{tab:wordshare}).

\subsubsection{Main Results}
Recall from Section 2.1 that we focus on three stock reaction indicators: log volatility $\log \sigma_{i,t}$, where $\sigma_{i,t}^2$ is defined in \eqref{eq:Volatility}, detrended log trading volume $V_{i,t}$ as in \eqref{eq:Volume} and returns $R_{i,t}$.  We first consider analyzing these three indicators with one trading day into the future, and use the following (separate) panel regressions.
\begin{eqnarray}
\log \sigma_{i, t+1} &=& \alpha + \beta_{1}I_{i, t} + \beta_{2}Pos_{i, t} + \beta_{3}Neg_{i, t} + \beta_{4}^\top X_{i,t} + \gamma_{i}+\varepsilon_{i,t},\\\label{eq:PanelVolatility}
V_{i, t+1} &=& \alpha + \beta_{1}I_{i, t} + \beta_{2}Pos_{i, t} + \beta_{3}Neg_{i, t} + \beta_{4}^\top X_{i,t} + \gamma_{i}+\varepsilon_{i,t},\\\label{eq:PanelVolume}
R_{i, t+1} &=& \alpha + \beta_{1}I_{i, t} + \beta_{2}Pos_{i, t} + \beta_{3}Neg_{i, t} + \beta_{4}^\top X_{i,t} + \gamma_{i}+\varepsilon_{i,t}.\label{eq:PanelReturns}
\end{eqnarray}
where $\gamma_i$ is the fixed effect for stock symbol $i$ satisfying $\sum_{i}\gamma_i=0$. $X_{i,t}$ is a vector of control variables that includes a set of market variables to control for systematic risk such as (1) S\&P 500 index return ($R_{M,t}$) to control for general market returns; (2) the CBOE VIX index on date $t$ to measure the generalized risk aversion ($VIX_{t}$); and a set of firm idiosyncratic variables such as (3) the lagged log volatility ($\log \sigma_{i,t}$); (4) the lagged return ($R_{i,t}$); (5) the lagged detrended log trading volume ($V_{i,t}$), where the lagged dependent variable is used to capture the persistence and omitted variables.    These three indicators essentially have a triple dynamic correlation, and they have been modeled as a trivariate vector autoregressive (VAR) model, see \cite{Chen&etal:2001} and \cite{Lee&Rui:2002}. Our indicators in Eqs.(\ref{eq:PanelVolatility}) to (\ref{eq:PanelReturns}) not only have themselves dynamic relationship with their lagged values, but also are impacted by the other lagged indicators.  We incorporate clustered standard errors by \cite{Arellano1987} as they allow for both time and cross-sectional dependence in the residuals. \cite{Petersen:2009} concludes that standard errors clustered on both dimensions are unbiased and achieve correctly sized confidence intervals while ordinary least squares standard errors might be biased in a panel data setting.

\begin{regtable5}{h!}
	\centering
	\scalebox{0.7}{
		\begin{tabular}{l|SSSSSSSS} \hline \hline
			Variable & \multicolumn{2}{c}{BL} & \multicolumn{2}{c}{LM} & \multicolumn{2}{c}{MPQA} & \multicolumn{2}{c}{PCA} \\\hline
			& \multicolumn{8}{c}{Panel A: Future Log Volatility $ \log \sigma_{i,t+1}$} \\
	   $I_{i,t}$ & -0.005 & (0.009) & -0.019$^{***}$ & (0.007) & -0.004 &
	   (0.010) & -0.014 & (0.010) \\
	   $Pos_{i,t}$ & -0.396$^{*}$ & (0.228) & 0.156 & (0.378) &
	   -0.517$^{**}$ & (0.217) & -0.210 & (0.201) \\
	   $Neg_{i,t}$ & 0.905$^{***}$ & (0.257) & 0.942$^{***}$ & (0.271) &
	   1.464$^{***}$ & (0.325) & 1.041$^{***}$ & (0.247) \\
	   $R_{M,t}$ & -1.507$^{***}$ & (0.217) & -1.501$^{***}$ & (0.216) &
	   -1.500$^{***}$ & (0.216) & -1.505$^{***}$ & (0.216) \\
	   $VIX_{t}$ & 2.329$^{***}$ & (0.085) & 2.335$^{***}$ & (0.085) &
	   2.331$^{***}$ & (0.086) & 2.330$^{***}$ & (0.085) \\
	   $\log \sigma_{i,t}$ & 0.242$^{***}$ & (0.010) & 0.242$^{***}$ & (0.010) &
	   0.242$^{***}$ & (0.010) & 0.242$^{***}$ & (0.010) \\
	   $R_{i,t}$ & 1.652$^{***}$ & (0.196) & 1.653$^{***}$ & (0.196) &
	   1.651$^{***}$ & (0.196) & 1.653$^{***}$ & (0.196) \\
	   $V_{i,t}$ & 0.065$^{***}$ & (0.006) & 0.065$^{***}$ & (0.006) &
	   0.065$^{***}$ & (0.006) & 0.065$^{***}$ & (0.006) \\
			\hline
			& \multicolumn{8}{c}{Panel B: Future Detrended Log Trading Volume $V_{i,t+1}$} \\
$I_{i,t}$ & 0.040$^{***}$ & (0.008) & 0.027$^{***}$ & (0.005) &
0.046$^{***}$ & (0.009) & 0.035$^{***}$ & (0.008) \\
$Pos_{i,t}$ & -0.496$^{***}$ & (0.188) & 0.051 & (0.275) &
-0.483$^{**}$ & (0.194) & -0.274$^{*}$ & (0.166) \\
$Neg_{i,t}$ & 0.726$^{***}$ & (0.257) & 0.563$^{**}$ & (0.251) &
0.548$^{*}$ & (0.290) & 0.590$^{**}$ & (0.232) \\
$R_{M,t}$ & -3.625$^{***}$ & (0.181) & -3.620$^{***}$ & (0.181) &
-3.617$^{***}$ & (0.181) & -3.622$^{***}$ & (0.181) \\
$VIX_{t}$ & -0.492$^{***}$ & (0.027) & -0.487$^{***}$ & (0.027) &
-0.487$^{***}$ & (0.027) & -0.489$^{***}$ & (0.027) \\
$\log \sigma_{i,t}$ & 0.132$^{***}$ & (0.004) & 0.132$^{***}$ & (0.004) &
0.132$^{***}$ & (0.004) & 0.132$^{***}$ & (0.004) \\
$R_{i,t}$ & 1.164$^{***}$ & (0.126) & 1.166$^{***}$ & (0.126) &
1.164$^{***}$ & (0.126) & 1.166$^{***}$ & (0.126) \\
			\hline
			& \multicolumn{8}{c}{Panel C: Future Returns $R_{i,t+1}$} \\
   $I_{i,t}$ & 0.000 & (0.000) & 0.000 & (0.000) & 0.000 & (0.000) &
   -0.000 & (0.000) \\
   $Pos_{i,t}$ & 0.019$^{***}$ & (0.007) & 0.030$^{***}$ & (0.011) &
   0.014$^{*}$ & (0.008) & 0.018$^{***}$ & (0.006) \\
   $Neg_{i,t}$ & -0.004 & (0.008) & -0.000 & (0.010) & -0.009 & (0.010)
   & -0.003 & (0.008) \\
   $R_{M,t}$ & -0.050$^{***}$ & (0.006) & -0.050$^{***}$ & (0.006) &
   -0.050$^{***}$ & (0.006) & -0.050$^{***}$ & (0.006) \\
   $VIX_{t}$ & 0.011$^{***}$ & (0.001) & 0.011$^{***}$ & (0.001) &
   0.011$^{***}$ & (0.001) & 0.011$^{***}$ & (0.001) \\
   $\log \sigma_{i,t}$ & -0.001$^{***}$ & (0.000) & -0.001$^{***}$ & (0.000)
   & -0.001$^{***}$ & (0.000) & -0.001$^{***}$ & (0.000) \\
   $R_{i,t}$ & -0.018$^{***}$ & (0.007) & -0.018$^{***}$ & (0.007) &
   -0.018$^{***}$ & (0.007) & -0.018$^{***}$ & (0.007) \\
   $V_{i,t}$ & 0.000 & (0.000) & 0.000 & (0.000) & 0.000 & (0.000) &
   0.000 & (0.000) \\
			\hline
			\hline
			
			\multicolumn{9}{l}{ \parbox{8in}{\vspace{3pt} $^{***}$ refers to a $p$ value less than 0.01, $^{**}$ refers to a $p$ value more than or equal to 0.01 and smaller than 0.05, and $^{*}$ refers to a $p$ value more than or equal to 0.05 and less than 0.1. Values in parentheses are clustered standard errors.}} \\
		\end{tabular}	
	}
	\vspace{-6pt}
	\caption{Entire Panel Regression Results}
	\label{tab:EntirePanel1}
\end{regtable5}
To answer our research question (i), if the three lexica are not consistent in their analytic ability to produce stock reaction indicators, we would expect that the value and the significance of $\beta_1$, $\beta_2$ or $\beta_3$ varies across three lexical projections.  For question (ii), if the positive and negative sentiments have asymmetric impacts, we would expect that $\beta_2$ and $\beta_3$ have different signs or significance. To address question (iii), we would expect that the value and the significance of $\beta_1$, $\beta_2$ or $\beta_3$ varies with different attention levels and in particular that the coefficient size is larger for higher attention firms.  As to question (iv), we would expect that the coefficients of sentiment variables are sector-specific.

We will discuss the analysis of different attention levels and different sectors respectively in Sections \ref{sect:att} and \ref{sect:sect}, and focus now on the entire sample.  The regression results are given in Table \ref{tab:EntirePanel1}. Results in Panel A indicate that the arrival of articles ($I_{i,t}$) distilled using the LM method is strongly negatively related to future log volatility, and that contingent on arriving articles, the negative sentiment distilled using the three methods is significantly positively related to future log volatility, whereas the positive sentiment distilled using the BL and MPQA methods is significantly negatived related to future log volatility. Results in Panel B show that contingent on arriving articles, the positive and negative sentiment have asymmetric strong impacts on future detrended log trading volume: the negative sentiment across three lexica strongly drives up future detrended log trading volume, whereas the positive sentiment distilled using the BL and MPQA methods is strongly negatively related to future detrended log trading volume.  The arrival of articles also strongly drives up future detrended log trading volume across three lexica.  These findings support the mixture of distribution hypothesis originated by \cite{Clark:1973}. As to future returns in Panel C, across three lexica and contingent on arriving articles, the positive sentiments are strongly positively related to future returns whereas the negative sentiment is unrelated to future returns. This finding sheds light on case against one unpleasant finding from \cite{Antweiler:2004} in which bullishness is not statistically significant for future return. It is interesting to note that the coefficients for the control variables do not vary much across lexical projections, which indicates that the sentiment measures are not so much correlated with the control variables and indeed provide incremental information.

It is difficult to diagnose a consensual performance from Table \ref{tab:EntirePanel1} because each lexicon may not fully reflect the complete sentiment and may have its own idiosyncratic nature as being evident from Table \ref{tab:worduniq}.  To overcome this problem that none of the lexica is perfectly complete, we design an artificial sentiment index: the first principal component, to capture a common component of three lexica and to take into account the fact from Figures \ref{fig:sent_pos} and \ref{fig:sent_neg} that they reveal the shared sentiment. The positive (negative) sentiment index explains 94.14\% (92.33\%) of the total sample variance.  As seen in the last column of Table \ref{tab:EntirePanel1}, these general positive and negative sentiment indices are beneficial to achieve more consistent and interpretable results. The negative sentiment index spurs the future stock volatility and trading volume. However, the positive sentiment index has very restrictive influence on future volatility, and suppresses the trading volume but increases stock returns.

\subsubsection{Sentiment Effect with Larger Lags and Neutral Sentiment}
Based on the sequential arrival of information hypothesis \citep[hereafter SAIH,][]{Copeland:1976,Copeland:1977}, information arrives to traders at different times and hence relationship with lags larger than one can exist. Hence, we extend the length of lag under investigation to be two to five trading days and run regressions using the entire sample. From Table \ref{tab:EntirePanel-noncum}, volatility still reacts to the news in lagged two days but no more earlier than it: lagged two day negative sentiment extracted by BL and LM are influential, indicating that the SAIH has been observed here but lagged relationship is restricted to past one and two day while article was posting. In this sense, the market seems efficient to incorporate information no longer than two days. Likewise, we find the negative sentiment in lagged two day still has an influence on future return. The coefficients across three lexicon projections are significant but positive. The coefficients of negative sentiment projected by the BL and LM methods are significant but positive. The negative sign, even insignificant, in lagged one day turns positive in lagged two day to reflect that stock returns revert to mean value, which is consistent with \cite{Antweiler:2004}.  Although not significant, the coefficients' sign for lag one indicates a slight negative influence on tomorrow's stock returns, but return will revert to its mean value in two days later shown by positive sign as negative news vanish. The sooner reversion is the more efficient market is. For the detrended log trading volume, the lagged effect is relatively insignificant.

Financial market is characterized by the clustering of information (news) arrival, so that we will see the volatility clustering \citep{Engle:2004}. The clustering of arrival of sentimental information motivates us to accumulate the sentiment variables from past trading days. Let $I_{i,t:(t+h-1)}$, $Pos_{i,t:(t+h-1)}$ and $Neg_{i,t:(t+h-1)}$ denote the indicator of arrival of articles, the average proportion of positive words and the average proportion of negative words based on articles published on or after trading day $t$ and before trading day $t+h$. Strikingly, the accumulated sentiment effect projected by BL and LM method on future volatility shown in Table \ref{tab:EntirePanel-cum} is very clear and keeps asymmetric, that is, only reacts to negative not to positive sentiment. Sometimes the sentiment news arrive consecutively and its accumulated influence lasts up to five trading days (one week). The accumulative sentiment effect can be also observed on the detrended log trading volume while accumulating to lagged four and five days, and on the future return while accumulating to lagged two days.

We also tried to consider the proportion of neutral words and examine its impact. Based on the neutral proportion defined by MPQA method, in general we find the neutral words have no influence in stock indicators. The results can be provided upon the request.

\FloatBarrier
\begin{regtable5}{ht}
	\scalebox{0.62}{
		\centering
		\begin{tabular}{l|SSSSSSSSS}
			\hline \hline
			& \multicolumn{3}{c}{BL}& \multicolumn{3}{c}{LM}& \multicolumn{3}{c}{MPQA} \\
			{Lag $h$} & {$I_{i,t}$} & {$Pos_{i,t}$} & {$Neg_{i,t}$} & {$I_{i,t}$} & {$Pos_{i,t}$} & {$Neg_{i,t}$} & {$I_{i,t}$} & {$Pos_{i,t}$} & {$Neg_{i,t}$} \\ \hline
			& \multicolumn{9}{c}{Panel A: Future Volatility $ \log \sigma_{i,t+h}$} \\
			$h=2$ & -0.000 & 0.000 & 0.005$^{*}$ & -0.000 & 0.001 & 0.005$^{*}$ & -0.000 & 0.001 & 0.004 \\
			& (0.000) & (0.002) & (0.003) & (0.000) & (0.003) & (0.003) & (0.000) & (0.002) & (0.003) \\
			$h=3$ & -0.000 & -0.001 & 0.003 & -0.000 & 0.001 & 0.004 & -0.000 & 0.000 & 0.003 \\
			& (0.000) & (0.002) & (0.003) & (0.000) & (0.003) & (0.003) & (0.000) & (0.002) & (0.003) \\
			$h=4$ & -0.000 & 0.000 & 0.002 & -0.000 & 0.002 & 0.004 & -0.000 & 0.001 & 0.000 \\
			& (0.000) & (0.002) & (0.003) & (0.000) & (0.003) & (0.003) & (0.000) & (0.002) & (0.003) \\
			$h=5$ & 0.000 & -0.002 & 0.002 & 0.000 & -0.002 & 0.003 & -0.000 & -0.000 & 0.001 \\
			& (0.000) & (0.002) & (0.003) & (0.000) & (0.003) & (0.003) & (0.000) & (0.002) & (0.003) \\
			\hline
			& \multicolumn{9}{c}{Panel B: Future Detrended Log Trading Volume $V_{i,t+h}$} \\
			$h=2$ & 0.003 & 0.112 & -0.198 & 0.004 & 0.079 & -0.158 & 0.003 & 0.006 & -0.414 \\
			& (0.006) & (0.140) & (0.174) & (0.005) & (0.227) & (0.183) & (0.007) & (0.140) & (0.219) \\
			$h=3$ & 0.001 & -0.011 & -0.082 & 0.001 & -0.003 & -0.125 & 0.002 & -0.170 & -0.188 \\
			& (0.006) & (0.140) & (0.174) & (0.005) & (0.227) & (0.183) & (0.007) & (0.140) & (0.219) \\
			$h=4$ & -0.001 & 0.064 & -0.539 & 0.004 & -0.324 & -0.556 & 0.001 & -0.020 & -0.811 \\
			& (0.006) & (0.140) & (0.488) & (0.005) & (0.227) & (0.536) & (0.007) & (0.140) & (0.479) \\
			$h=5$ & 0.008 & -0.208 & -0.410 & -0.004 & -0.022 & -0.096 & 0.001 & -0.069 & -0.416 \\
			& (0.006) & (0.140) & (0.301) & (0.005) & (0.227) & (0.183) & (0.007) & (0.140) & (0.278) \\
			\hline
			& \multicolumn{9}{c}{Panel C: Future Returns $R_{i,t+h}$} \\
			$h=2$ & -0.000 & 0.000 & 0.016$^{*}$ & -0.000 & -0.003 & 0.024$^{**}$ & -0.000 & 0.001 & 0.026$^{**}$ \\
			& (0.000) & (0.007) & (0.009) & (0.000) & (0.012) & (0.010) & (0.000) & (0.008) & (0.012) \\
			$h=3$ & 0.000 & -0.001 & -0.001 & 0.000 & -0.010 & 0.005 & 0.001 & -0.011 & 0.003 \\
			& (0.000) & (0.008) & (0.009) & (0.000) & (0.012) & (0.010) & (0.000) & (0.008) & (0.012) \\
			$h=4$ & -0.000 & 0.001 & 0.016$^{*}$ & -0.000 & 0.010 & 0.006 & -0.000 & -0.003 & 0.011 \\
			& (0.000) & (0.007) & (0.009) & (0.000) & (0.012) & (0.010) & (0.000) & (0.008) & (0.012) \\
			$h=5$& 0.000 & -0.011 & 0.009 & 0.000 & -0.018 & 0.002 & 0.000 & -0.013 & 0.014 \\
			& (0.000) & (0.007) & (0.009) & (0.000) & (0.012) & (0.010) & (0.000) & (0.009) & (0.012) \\
			\hline \hline
			\multicolumn{10}{l}{\parbox{9.5in}{\vspace{4pt} $^{***}$ refers to a $p$ value less than 0.01, $^{**}$ refers to a $p$ value more than or equal to 0.01 and smaller than 0.05, and $^{*}$ refers to a $p$ value more than or equal to 0.05 and less than 0.1. Values in parentheses are standard errors.}} \\
			\end{tabular}
			}
			\vspace{-6pt}
			\caption{Entire Panel Regression Results with Larger Lags (Noncumulative Articles)}
			\label{tab:EntirePanel-noncum}
			\end{regtable5}
			\FloatBarrier
			
			\FloatBarrier
			\begin{regtable5}{ht}
				\scalebox{0.62}{
					\centering
					\begin{tabular}{l|SSSSSSSSS}
						\hline \hline
						& \multicolumn{3}{c}{BL}& \multicolumn{3}{c}{LM}& \multicolumn{3}{c}{MPQA} \\
						{Lag $h$} & {$I_{i,t:(t+h-1)}$} & {$Pos_{i,t:(t+h-1)}$} & {$Neg_{i,t:(t+h-1)}$} & {$I_{i,t:(t+h-1)}$} & {$Pos_{i,t:(t+h-1)}$} & {$Neg_{i,t:(t+h-1)}$}  & {$I_{i,t:(t+h-1)}$} & {$Pos_{i,t:(t+h-1)}$} & {$Neg_{i,t:(t+h-1)}$}  \\ \hline
						& \multicolumn{9}{c}{Panel A: Future Volatility $ \log \sigma_{i,t+h}$} \\
						$h=2$ & -0.000 & -0.001 & 0.006$^{**}$ & -0.000 & -0.001 & 0.007$^{**}$ & -0.000 & -0.001 & 0.004 \\
						& (0.000) & (0.002) & (0.002) & (0.000) & (0.003) & (0.003) & (0.000) & (0.002) & (0.003) \\
						$h=3$ & -0.000 & -0.002 & 0.006$^{***}$ & -0.000$^{*}$ & -0.001 & 0.008$^{***}$ & -0.000 & -0.001 & 0.005 \\
						& (0.000) & (0.002) & (0.002) & (0.000) & (0.003) & (0.003) & (0.000) & (0.002) & (0.003) \\
						$h=4$ & -0.000 & -0.001 & 0.006$^{***}$ & -0.000$^{**}$ & -0.000 & 0.008$^{***}$ & -0.000 & -0.001 & 0.003 \\
						& (0.000) & (0.002) & (0.002) & (0.000) & (0.003) & (0.003) & (0.000) & (0.002) & (0.003) \\
						$h=5$ & 0.000 & -0.003 & 0.006$^{**}$ & -0.000 & -0.002 & 0.008$^{***}$ & -0.000 & -0.002 & 0.003 \\
						& (0.000) & (0.002) & (0.002) & (0.000) & (0.003) & (0.003) & (0.000) & (0.002) & (0.003) \\
						\hline
						& \multicolumn{9}{c}{Panel B: Future Detrended Log Trading Volume $V_{i,t+h}$} \\
						$h=2$ & 0.006 & -0.016 & -0.133 & 0.008$^{*}$ & -0.148 & -0.187 & 0.002 & 0.006 & -0.253 \\
						& (0.006) & (0.125) & (0.156) & (0.004) & (0.203) & (0.169) & (0.006) & (0.126) & (0.198) \\
						$h=3$ & 0.006 & -0.072 & -0.111 & 0.005 & -0.189 & -0.078 & 0.001 & -0.063 & -0.174 \\
						& (0.005) & (0.123) & (0.153) & (0.004) & (0.198) & (0.167) & (0.006) & (0.124) & (0.193) \\
						$h=4$ & 0.008 & -0.310$^{**}$ & -0.096 & 0.010$^{**}$ & -0.486$^{**}$ & -0.293$^{*}$ & 0.004 & -0.138 & -0.473 \\
						& (0.005) & (0.152) & (0.124) & (0.004) & (0.200) & (0.168) & (0.006) & (0.125) & (0.327) \\
						$h=5$ & 0.014${**}$ & -0.242$^{*}$ & -0.408$^{*}$ & 0.008$^{*}$ & -0.428$^{**}$ & -0.228 & 0.009 & -0.193 & -0.646 \\
						& (0.006) & (0.126) & (0.246) & (0.004) & (0.202) & (0.171) & (0.007) & (0.126) & (0.493) \\
						\hline
						& \multicolumn{9}{c}{Panel C: Future Returns $R_{i,t+h}$} \\
						$h=2$ & -0.001$^{*}$ & 0.013$^{**}$ & 0.009 & -0.000 & 0.019$^{*}$ & 0.010 & -0.001$^{**}$ & 0.013$^{*}$ & 0.009 \\
						& (0.000) & (0.007) & (0.008) & (0.000) & (0.011) & (0.009) & (0.000) & (0.007) & (0.010) \\
						$h=3$ & -0.000 & 0.009 & 0.004 & -0.000 & 0.013 & 0.007 & -0.000 & 0.004 & 0.007 \\
						& (0.000) & (0.007) & (0.008) & (0.000) & (0.011) & (0.009) & (0.000) & (0.007) & (0.010) \\
						$h=4$ & -0.000 & 0.008 & 0.012 & -0.000 & 0.017 & 0.009 & -0.001 & 0.005 & 0.016 \\
						& (0.000) & (0.007) & (0.008) & (0.000) & (0.011) & (0.009) & (0.000) & (0.007) & (0.010) \\
						$h=5$& -0.000 & 0.000 & 0.010 & -0.000 & 0.004 & 0.006 & -0.000 & -0.003 & 0.019 \\
						& (0.000) & (0.007) & (0.008) & (0.000) & (0.011) & (0.009) & (0.000) & (0.007) & (0.010) \\
						\hline \hline
						\multicolumn{10}{l}{\parbox{9.5in}{\vspace{4pt} $^{***}$ refers to a $p$ value less than 0.01, $^{**}$ refers to a $p$ value more than or equal to 0.01 and smaller than 0.05, and $^{*}$ refers to a $p$ value more than or equal to 0.05 and less than 0.1. Values in parentheses are standard errors.}} \\
						\end{tabular}
						}
						\vspace{-6pt}
						\caption{Entire Panel Regression Results with Larger Lags (Cumulative Articles)}
						\label{tab:EntirePanel-cum}
						\end{regtable5}
						\FloatBarrier

\subsubsection{Monte Carlo Simulation based on Entire Sample Results}
\label{sect:MonteCarloSim}
\vspace{-6pt}

The text sentiment effects, as reported in Table \ref{tab:EntirePanel1}, allow us deeper insights and analysis.  More precisely we may address the important question of asymmetric reactions to the given sentiment scales.  In order to do so we employ Monte Carlo techniques to investigate different facets of the sentiment effects.  The components of this Monte Carlo study are: (1) to simulate the appearance of articles with presumed probabilities; (2) to provide a realistic set of scenarios regarding the frequency and content (positive v.s. negative) of articles;  (3) to obtain volatility induced by the generated article (using Table \ref{tab:EntirePanel1}); (4) to demonstrate the impact of synthetic text on future volatility;  (5) to visualize and test an asymmetry effect as formulated in research question (ii).

The simulation scenarios (for each variable involved) are summarized briefly as follows.  We employ a Bernoulli random variable $I_{i,t}$ indicating that articles arrive at a specific frequency $p_i$, where for each individual stock symbol $i$, $p_i$ is estimated by the fraction of days with at least one relevant article. Given the outcome of this article indicator, we generate the corresponding positive and negative proportions through a copula approach using the conditional inversion method as described in \cite{Frees:1998}. We follow the two-step approach that is widely mentioned in literature such as \cite{Patton:2006}, \cite{Hotta:2006} and \cite{DiClemente:2004}. In the first step, the marginal distributions are modeled by their corresponding empirical distribution function (edf) to avoid imposing a parametric distribution; in the second step, a Gaussian copula is estimated to take the inherent dependence among variables into account.  For the sentiment variables, this approach is applied to each firm separately since each firm has a different $p_i$ and only days with at least one article relevant to the firm are included in the estimation. To simulate market returns $R_{M,t}$ and individual returns $R_{i,t}$ for all 100 symbols, we first filter these variables by estimated MA(1)-GARCH(1,1) processes and standardize the residuals by dividing them by estimated standard deviations. We then apply the copula approach to the standardized residuals, and the simulated standardized residuals are transformed into simulated values of $R_{M,t}$ or $R_{i,t}$ by multiplying them by the median of the priorly estimated standard deviations for the market or the specific firm $i$.  The company specific fixed effects $\gamma_i$ are not incorporated as the simulated volatility for different firms is otherwise not graphically comparable.  For the other control variables, CBOE VIX index $VIX_t$ is fixed at its mean value over the sample period, and past log volatility and past detrended log trading volume are not used in the simulation.

\begin{figure}[h] \centering
\begin{tikzpicture}[
        every node/.style={anchor=south west,inner sep=0pt},
        x=1mm, y=1mm,]
     \node (fig1) at (0,0)
       {\includegraphics[width = 6.3in]{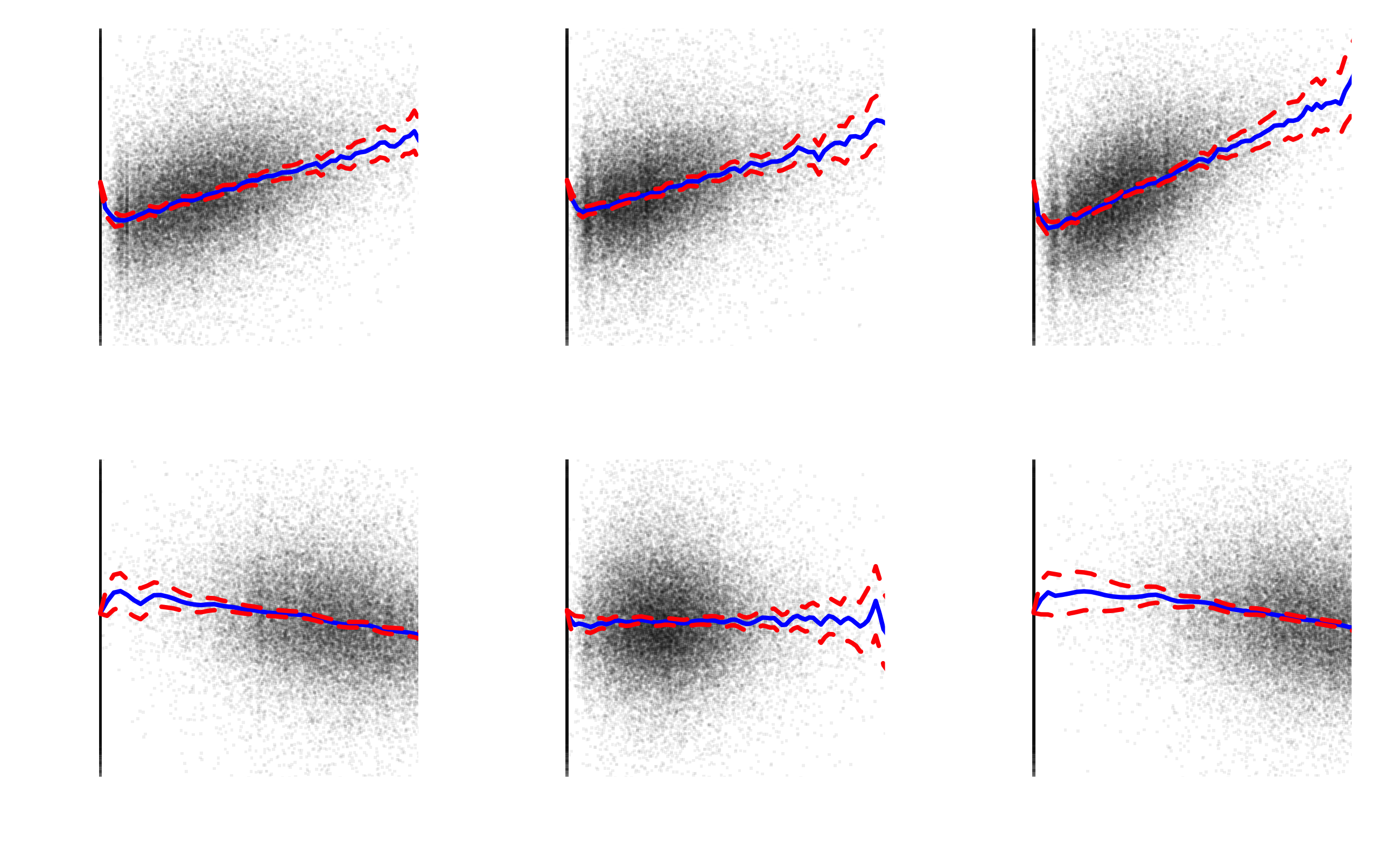}};
     \node (fig2) at (0,0)
       {\includegraphics[width = 6.3in]{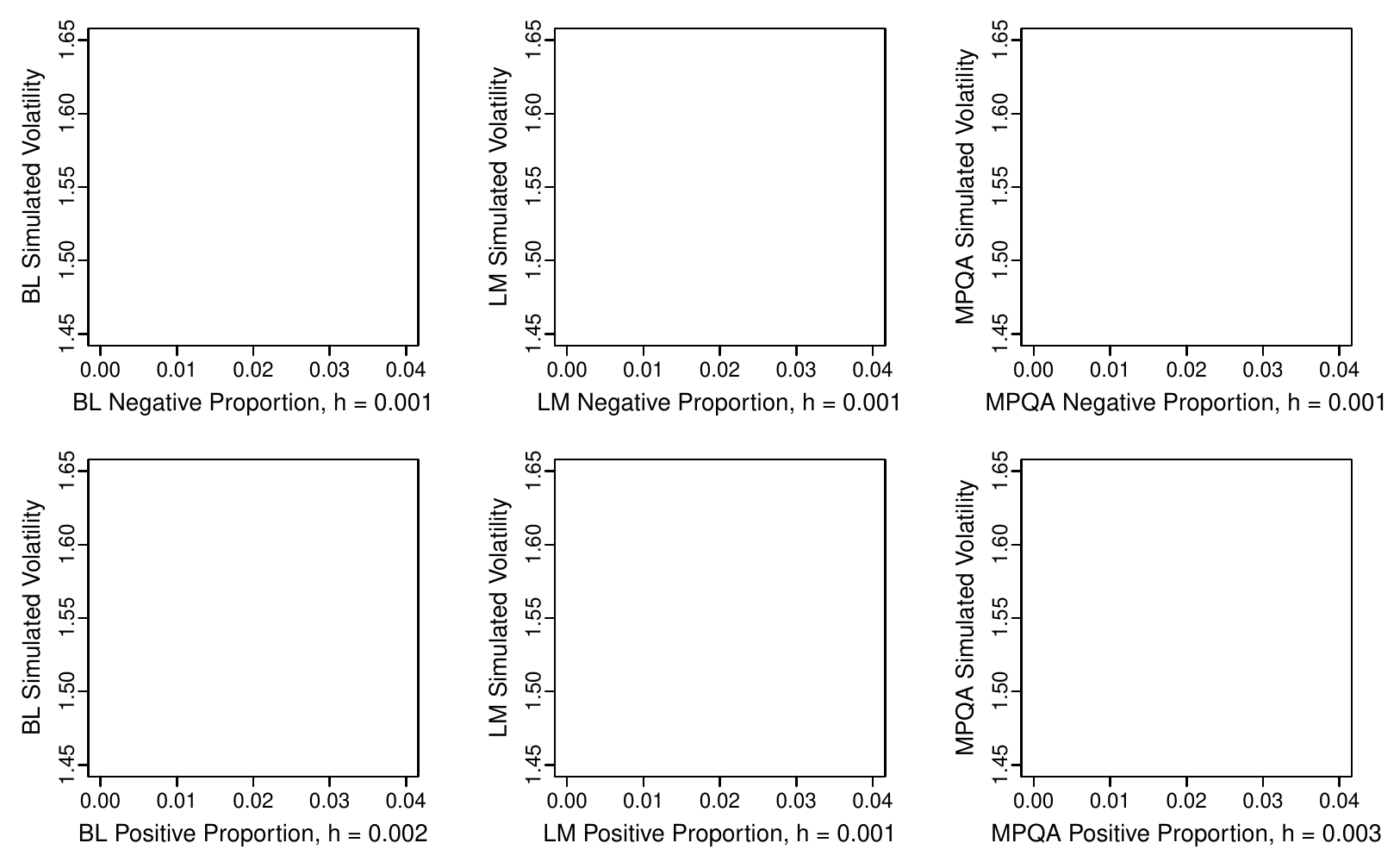}};
\end{tikzpicture}
\vspace{-6pt}
\caption{Monte Carlo Simulation based on Entire Sample Results}
\label{fig:entire_panel_scat}
\end{figure}

Figure \ref{fig:entire_panel_scat} demonstrates, for one simulation, the association between the negative and positive proportions as distilled via our three projection methods and their simulated future volatility outcomes. We estimate a local linear regression model (solid line) and corresponding 95\% uniform confidence bands based on \cite{Sun:1994}. Both are estimated using Locfit by \cite{Loader:1999} in the R environment. \cite{loader:1997} discuss the robustness of this approach and conclude that the results are conservative but reasonable for heavy tailed error distributions. The bandwidth is automatically chosen by using the plug-in selector according to \cite{ruppert:1995}. We limit the the visible area to sentiment values between $0$ and $0.04$ as well as volatility values between $1.45$ and $1.65$ to make the different lexica visually comparable. Nevertheless, all simulated values are utilized in the estimation of the regression curve and confidence bands. The clustered points lying on the vertical axis indicate that there is absence of articles. The range of this cluster from $0.77$ to $2.57$ is caused by the impact from the simulated control variables as well as the idiosyncratic impact captured by the residual term.

Apparently, an asymmetry effect becomes visible. One observes that the slopes of the volatility curves given negative sentiment is mainly positive while the curves for positive sentiment seem to be rather flat and even go down in the case of BL and MPQA methods. One can also compare the confidence bands to address the question whether negative sentiment has a significantly higher effect on the volatility than positive sentiment. The confidence bands of $Pos$ and $Neg$ do not overlap for sentiment values between $0.023$ and $0.056$ for BL, between $0.017$ and $0.039$ for LM and between $0.023$ and $0.05$ for MPQA.

This asymmetry effect parallels the well known imbalance of future volatility given good v.s. bad news. The leverage effect depicts a negative relation between the lagged return and the risk resulting from bad news that causes higher volatility. \cite{Black:1976} and \cite{Christie:1982} find that bad news in the financial market produce such an asymmetric effect on future volatility relative to good news. This leverage effect has also been shown by \cite{Bekaert&Wu:2000} and \cite{Feunou&Tedongap:2012}. In the same vein, \cite{Glosten&etal:1993} introduce GARCH with differing effects of negative and positive shocks taking into account the leverage effect.

\subsection{Does Attention Ratio matter?}
\label{sect:att}
While people post their text to express their opinions, or the comments to other articles, they are undoubtedly paying attention to the firm mentioned by their articles. In this respect article posting is a revealed attention measure.  In fact, in our collected 43,459 articles across 100 stocks, it is obvious that not every firm shares the attention equivalently. To reflect these differences, we define the attention ratio for a symbol as the number of days with articles divided by the total number of days in the sample period, 1,255.  The symbol ``AAPL" (Apple Computer Inc.) attracts the most attention with an attention ratio of 0.818. Articles involving AAPL arrive in social media almost every day (81.8 days over 100 days). However, the symbol ``TRV" (Travelers Companies, Inc.) has the lowest attention ratio, 0.204, which means that one finds a related article every five trading days, i.e. one week. Different from the ``indirect" attention measures from stock indicators such as trading volumes, extreme returns or price limits, this attention measure is a kind of ``direct" measure of investor attention, and shares the same idea as the Search Volume Index (SVI) constructed by Google. Beyond the SVI, our attention can be further projected to ``Positive" or ``Negative" attention. In our main research question (ii), we are interested in whether the well known asymmetric response (bad vs. good news) is appropriately reflected in the lexical projections. Assuming that investors are more risk-averse, they should be more aware of negative articles and pay more attention to them.

Attention is one of the basic elements in traditional asset pricing models. The conventional asset pricing models assume that information is instantaneously incorporated into asset prices when it arrives. The basic assumption behind this argument is that investors pay ``sufficient" attention to the asset. Under this condition, the market price of asset should be very efficient in incorporating any relevant news. In this aspect, the high attention firms should be more responsive to the text sentiment distilled from the articles, and their market prices should reflect this efficiency. As such, the high attention samples stand on the side of the traditional asset pricing models, and the findings from them are expected to support the efficient market hypothesis. However, attention in reality is a scarce cognitive resource, and investors have limited attention instead \citep{Kahneman:1973}.  Further research on this topic from \cite{Merton:1987}, \cite{Sims:2003} and \cite{Peng&Xiong:2006} confirms that the limited attention can affect asset pricing. The low attention firms with very limited attention may ineffectively or insufficiently reflect the text sentiment information, so that their corresponding stock reactions could be greatly bounded. This argument is in accordance with the fact that the limited attention causes stock prices to deviate from the fundamental values \citep{Hong&Stein:1999}, implying a potential arbitrage opportunity.

\subsubsection{Descriptive Statistics for the Firms with different Attention Ratios}
Grouping the samples by their attention ratios and examining the responses from different attention groups may offer a clue to the aforementioned conjectures. The criterion used to group the sample firms is based on the quantiles of the attention ratio.  Firms whose attention ratios are above the 75\% quantile (0.3693) are grouped as ``extremely high", between 50\% (0.3026) and 75\% quantiles as ``high", between 25\% (0.2455) and 50\% quantiles as ``median", and lower than 25\% quantile as ``low".  For each attention group, Table \ref{tab:DescriptiveFirmAR} reports across lexical projections the mean values of positive ($\mu_{Pos}$) and negative ($\mu_{Neg}$) sentiment proportions, calculated by averaging $Pos_{i,t}$ or $Neg_{i,t}$ over all relevant symbol-day combinations, the proportion of relevant symbol-day combinations with $Neg_{i,t}>Pos_{i,t}$, the average attention ratio, and the average number of days with articles, calculated by averaging the number of days with articles over all relevant symbols. The ``extreme high" groups receive an average attention ratio of 55.14\%, indicating on average these firms have been looked at every two days.  By contrast, the low attention group with an average attention ratio of 21.97\% receives attention at weekly frequency (5 trading days). By comparing the magnitude of $\mu_{Neg}$, one observes that investors are inclined to express negative sentiments in the ``extreme high" group. One may conclude therefore that higher attention is coming with a ``negative text", or inversely speaking: the negative article creates higher attention.  This is evident for example in the case of the LM method, where the proportion of symbol-day combinations with dominance of negative sentiment is 46\% in the ``extremely high" group.  For the constituents in this particular attention group, we find on average 691 days with articles observed over a total of 1255 sample days (5 years), which is almost three times the average number of days with articles for the low attention group.

\FloatBarrier
\begin{summarytable}{h}
	\scalebox{0.65}{
		\centering
	\begin{tabular}{l|S[table-space-text-post = 2]S[table-space-text-post = 2]S[table-space-text-post = 2]S[table-space-text-post = 2]S[table-space-text-post = 2]S[table-space-text-post = 2]S[table-space-text-post = 2]S[table-space-text-post = 2]S[table-space-text-post = 2]S[table-space-text-post = 2]c} \hline \hline
		& \multicolumn{3}{c}{BL} & \multicolumn{3}{c}{LM} & \multicolumn{3}{c}{MPQA} & {Attention} & {Number of Days} \\
		Attention & {$\mu_{Pos}$} & {$\mu_{Neg}$} & {$Neg>Pos$} & {$\mu_{Pos}$} & {$\mu_{Neg}$} & {$Neg>Pos$} & {$\mu_{Pos}$} & {$\mu_{Neg}$} & {$Neg>Pos$} & {Ratio} & {with Articles}\\ \hline
		Extremely high & 0.032   & 0.016   & 0.119    & 0.013   & 0.014   & 0.460    & 0.038     & 0.013     & 0.027      & 0.551 & 691 \\
		High           & 0.032   & 0.015   & 0.113    & 0.013   & 0.012   & 0.403    & 0.038     & 0.013     & 0.031      & 0.343 & 430 \\
		Median         & 0.035   & 0.014   & 0.083    & 0.014   & 0.011   & 0.339    & 0.039     & 0.012     & 0.027      & 0.273 & 356  \\
		Low            & 0.036   & 0.014   & 0.086    & 0.015   & 0.011   & 0.333    & 0.040     & 0.012     & 0.031      & 0.220 & 264  \\
		\hline \hline
	\end{tabular}
}
\caption{The Summary Statistics for different Attention Ratio Groups}
\label{tab:DescriptiveFirmAR}
\end{summarytable}
\FloatBarrier

\subsubsection{The Results of Attention Analysis}

The central interest of this research focuses on understanding to which extent distilled news flow and its derived parameters (like attention) impacts the relation between text sentiment and stock reactions. We employ panel regression designed for the given attention groups, and therefore each panel regression equally comprises of 25 sample firms. The results are displayed in Table \ref{tab:AttentionAnalysis}.  For the ``extremely high" group, the text sentiment carries a major and highly significant influence on future volatility consistently across the three lexical projections.  As a caveat though please note that the sentiment effect on volatility shown in Panel A is exclusive for negative news contingent on arriving articles, the stock volatility rarely reacts to positive or optimistic news. Panel B summarizes the attention analysis on the detrended log trading volume.
For the ``extremely high" group, in the LM and MPQA projection methods, arrival of articles ($I_{i,t}$) brings relevant information, and creates a growing trading volume, especially when it comes with negative news. The corresponding analysis for stock returns are also reasonable. The stock returns of ``high" group react clearly to the sentiments, contingent on arriving articles, they rise for optimistic news and decline for pessimistic consensus. In the case of LM method, the significant positive coefficient of $Neg_{i,t}$ for the ``extremely high" group suggests that the market participants act according to the uncertain market hypothesis developed by \cite{brown:1988} and based on the overreaction hypothesis by \cite{bondt:1985}. Here, the market participants set new prices before the full range of the news content is resolved. In case of unfavorable news, the investors set stock prices significantly below their conditional expected values and thus, react risk-averse. On the subsequent day, the mispriced stock price will revert to its true value.

The collected empirical evidence so far suggests that the distilled news of high attention firms effectively drive their stock volatilities, trading volumes and returns. They are highly responsive to the sentiment across lexical projections.

\FloatBarrier
\begin{regtable5}{h!}
	\scalebox{0.62}{
		\centering
		\begin{tabular}{l|SSSSSSSSS}
			\hline \hline
			& \multicolumn{3}{c}{BL}& \multicolumn{3}{c}{LM}& \multicolumn{3}{c}{MPQA} \\
			{Attention} & {$I_{i,t}$} & {$Pos_{i,t}$} & {$Neg_{i,t}$} & {$I_{i,t}$} & {$Pos_{i,t}$} & {$Neg_{i,t}$} & {$I_{i,t}$} & {$Pos_{i,t}$} & {$Neg_{i,t}$} \\ \hline
			& \multicolumn{9}{c}{Panel A: Future Volatility $ \log \sigma_{i,t+1}$} \\
			Low & 0.020 & -0.736 & -0.074 & 0.010 & -1.027 & -0.195 & 0.016 & -0.655 & 0.275 \\
			& (0.025) & (0.666) & (0.766) & (0.016) & (1.027) & (0.788) & (0.029) & (0.633) & (0.866) \\
			Median & 0.004 & -0.690 & 1.107$^{**}$ & -0.012 & -0.308 & 1.126$^{*}$ & 0.008 & -0.872$^{*}$ & 1.767$^{**}$ \\
			& (0.016) & (0.449) & (0.446) & (0.016) & (0.778) & (0.630) & (0.019) & (0.515) & (0.707) \\
			High & -0.016 & -0.460 & 1.324$^{***}$ & -0.046$^{***}$ & 0.967 & 1.806$^{***}$ & -0.019 & -0.636$^{**}$ & 2.548$^{***}$ \\
			& (0.017) & (0.442) & (0.475) & (0.013) & (0.724) & (0.615) & (0.016) & (0.315) & (0.662) \\
			Extremely & -0.010 & 0.027 & 0.784$^{**}$ & -0.013 & 0.483 & 0.747$^{**}$ & -0.002 & -0.182 & 0.909$^{**}$ \\
			High& (0.014) & (0.257) & (0.371) & (0.013) & (0.457) & (0.300) & (0.017) & (0.284) & (0.433) \\
			\hline
			& \multicolumn{9}{c}{Panel B: Future Detrended Log Trading Volume $V_{i,t+1}$} \\
			Low & 0.054$^{**}$ & -0.817 & 0.312 & 0.044$^{***}$ & -0.923 & -0.109 & 0.049$^{*}$ & -0.433 & -0.197 \\
			& (0.024) & (0.502) & (0.665) & (0.014) & (0.657) & (0.556) & (0.029) & (0.567) & (0.796) \\
			Median & 0.052$^{***}$ & -0.851$^{**}$ & 1.116$^{*}$ & 0.032$^{***}$ & -0.199 & 0.861 & 0.062$^{***}$ & -0.754$^{**}$ & 0.449 \\
			& (0.014) & (0.398) & (0.600) & (0.010) & (0.535) & (0.601) & (0.013) & (0.342) & (0.689) \\
			High & 0.036$^{***}$ & -0.198 & 0.554 & 0.021$^{*}$ & 0.815$^{*}$ & 0.447 & 0.046$^{***}$ & -0.358 & 0.419 \\
			& (0.009) & (0.299) & (0.459) & (0.011) & (0.487) & (0.451) & (0.016) & (0.385) & (0.559) \\
			Extremely & 0.023 & -0.242 & 0.958$^{**}$ & 0.017$^{*}$ & 0.299 & 0.796$^{*}$ & 0.032$^{**}$ & -0.408 & 1.084$^{**}$ \\
			High& (0.014) & (0.336) & (0.416) & (0.008) & (0.521) & (0.429) & (0.014) & (0.299) & (0.427) \\
			\hline
			& \multicolumn{9}{c}{Panel C: Future Returns $R_{i,t+1}$} \\
			Low & 0.000 & 0.012 & 0.009 & 0.000 & 0.021 & -0.001 & 0.000 & 0.010 & -0.016 \\
			& (0.001) & (0.022) & (0.023) & (0.000) & (0.030) & (0.023) & (0.001) & (0.021) & (0.032) \\
			Median & -0.001 & 0.024$^{*}$ & 0.009 & 0.000 & 0.035$^{*}$ & -0.022 & -0.001 & 0.034$^{*}$ & 0.007 \\
			& (0.001) & (0.012) & (0.018) & (0.000) & (0.019) & (0.024) & (0.001) & (0.018) & (0.024) \\
			High & 0.000 & 0.028$^{**}$ & -0.034$^{***}$ & 0.001$^{**}$ & 0.038$^{*}$ & -0.046$^{**}$ & 0.000 & 0.024$^{**}$ & -0.044$^{***}$ \\
			& (0.000) & (0.012) & (0.011) & (0.000) & (0.022) & (0.018) & (0.001) & (0.011) & (0.016) \\
			Extremely & 0.000 & 0.017 & 0.004 & -0.000 & 0.031 & 0.033$^{**}$ & 0.001$^{*}$ & -0.006 & 0.009 \\
			High& (0.000) & (0.012) & (0.012) & (0.000) & (0.021) & (0.013) & (0.000) & (0.011) & (0.016) \\
			\hline \hline
			\multicolumn{10}{l}{\parbox{9.5in}{\vspace{4pt} $^{***}$ refers to a $p$ value less than 0.01, $^{**}$ refers to a $p$ value more than or equal to 0.01 and smaller than 0.05, and $^{*}$ refers to a $p$ value more than or equal to 0.05 and less than 0.1. Values in parentheses are clustered standard errors.}} \\
		\end{tabular}
	}
	\vspace{-6pt}
	\caption{Attention Analysis: The Impact on future Volatility, Trading Volume and Returns}
	\label{tab:AttentionAnalysis}
\end{regtable5}
\FloatBarrier

Given the high attention received, any relevant information including the articles made by individual traders has been fully incorporated into their asset prices and dynamics. Due to their efficiency, the article posting and discussing today can predict stock reactions tomorrow.  For lower attention firms, one cannot make such a strong claim. Investors may think those firms are negligible and may therefore underreact to the available information. The underreaction from limited attention is likely to cause stock prices to deviate from the fundamental values, and an arbitrage opportunity may emerge. Our evidence is in line with \cite{Da&etal:2011} in which they support the attention-induced price pressure hypothesis. By using the SVI from Google as attention measure, they find stronger attention-induced price pressure among stocks in which individual investor attention matters most. Beyond their study, we find that high attention is usually accompanied with negative articles, and negative articles contribute more to attention and cause more stock reactions, supporting an asymmetric response.

It is interesting to note that the coefficients for the control variables do not vary much across lexical projections in each attention group (results not shown here), which indicates that for each attention group, the sentiment measures are not so much correlated with the control variables and provide incremental information.

\subsubsection{Monte Carlo Simulation based on Attention Analysis}

\begin{figure}[h!] \centering
\begin{tikzpicture}[
        every node/.style={anchor=south west,inner sep=0pt},
        x=1mm, y=1mm,]
     \node (fig1) at (0,0)
       {\includegraphics[width = 6.4in]{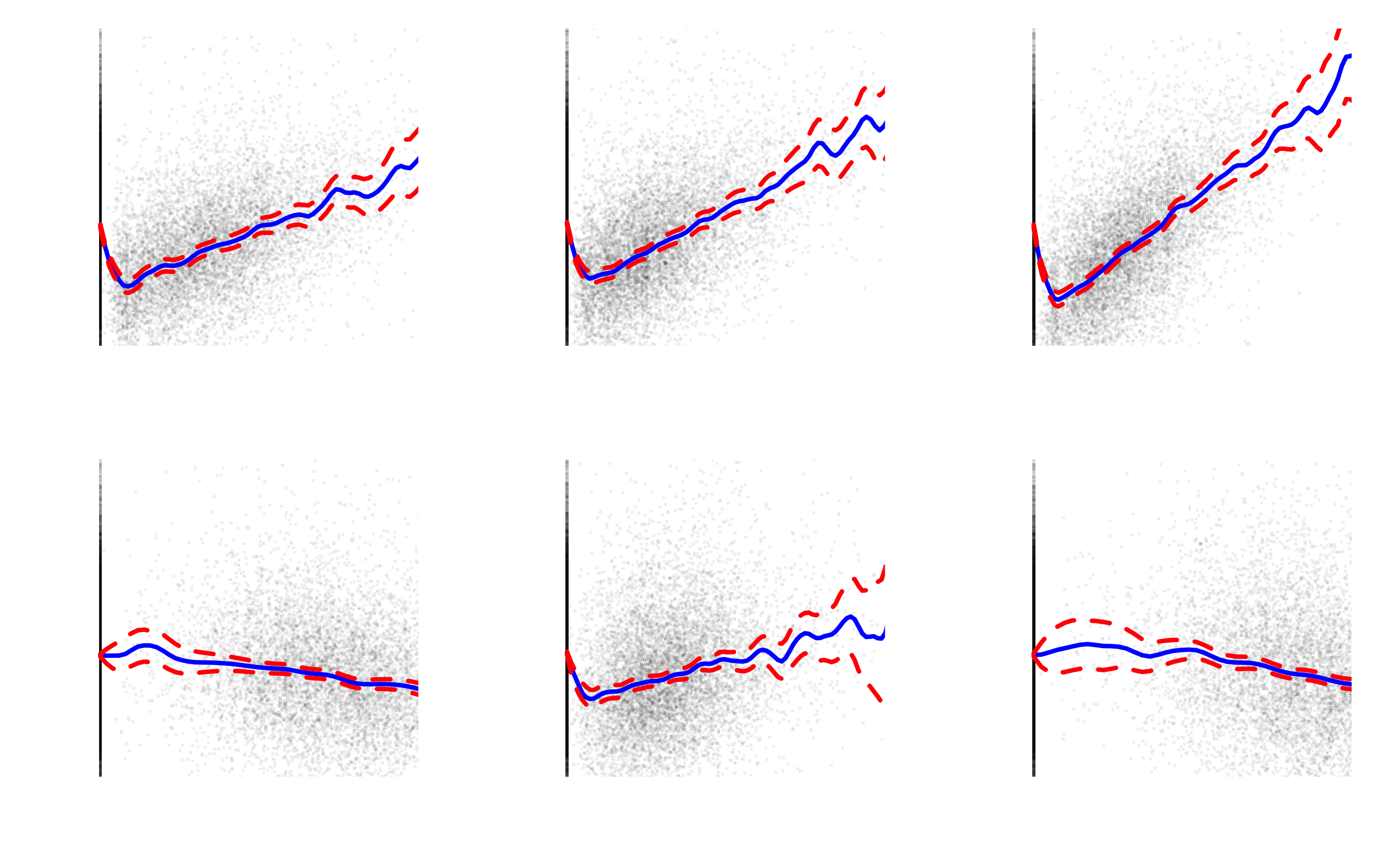}};
     \node (fig2) at (0,0)
       {\includegraphics[width = 6.4in]{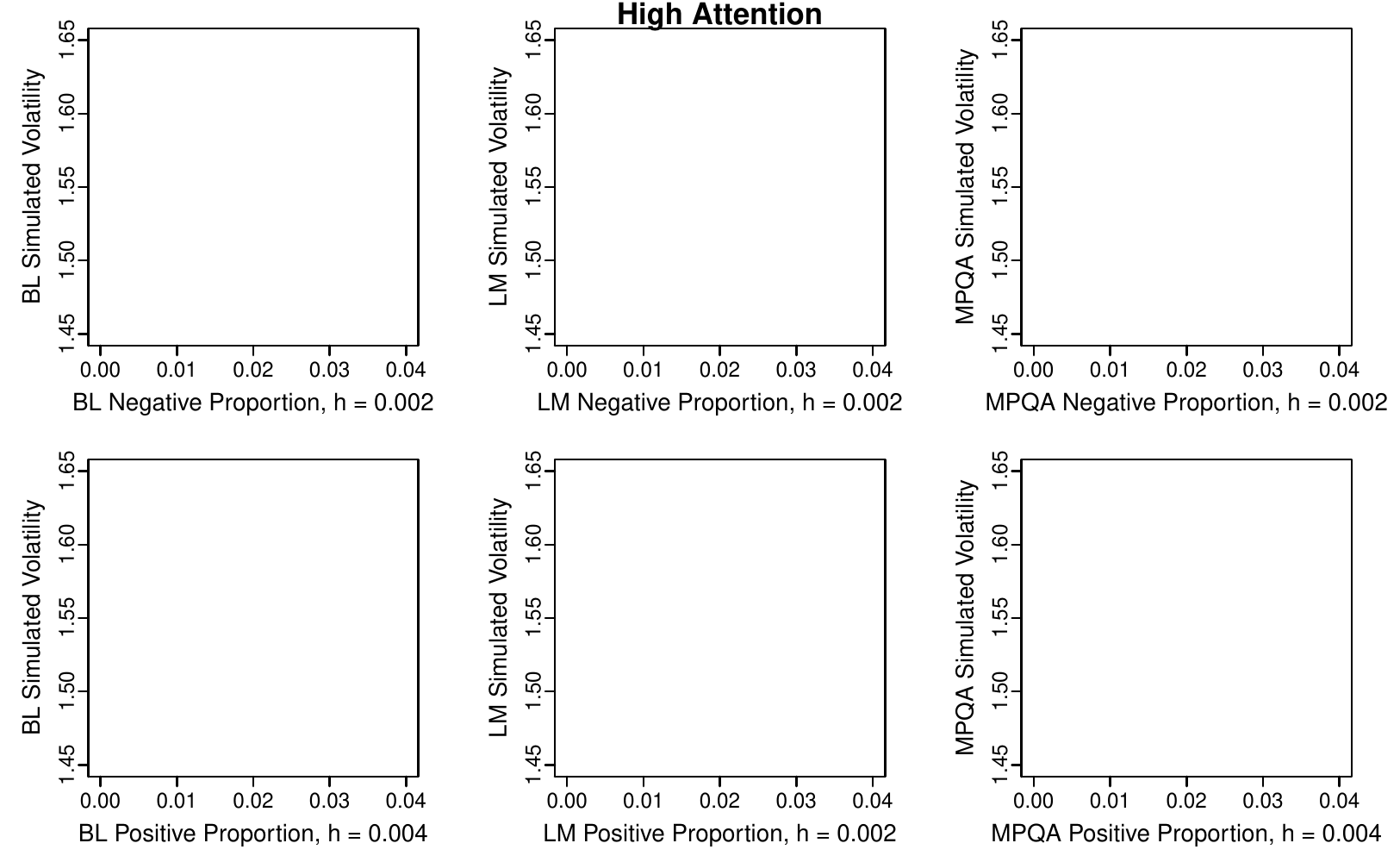}};
\end{tikzpicture}
\begin{tikzpicture}[
        every node/.style={anchor=south west,inner sep=0pt},
        x=1mm, y=1mm,]
     \node (fig1) at (0,0)
       {\includegraphics[width = 6.4in]{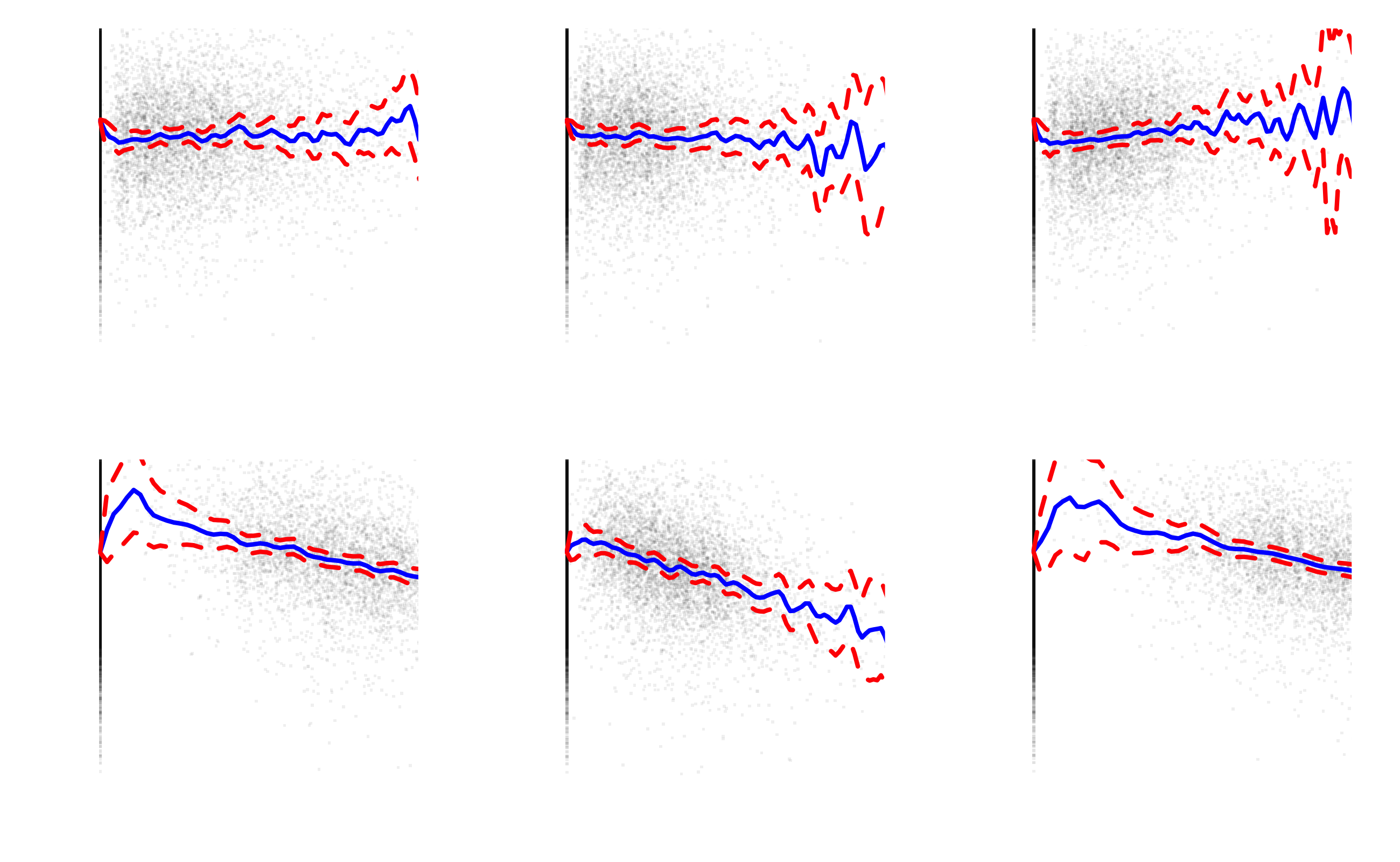}};
     \node (fig2) at (0,0)
       {\includegraphics[width = 6.4in]{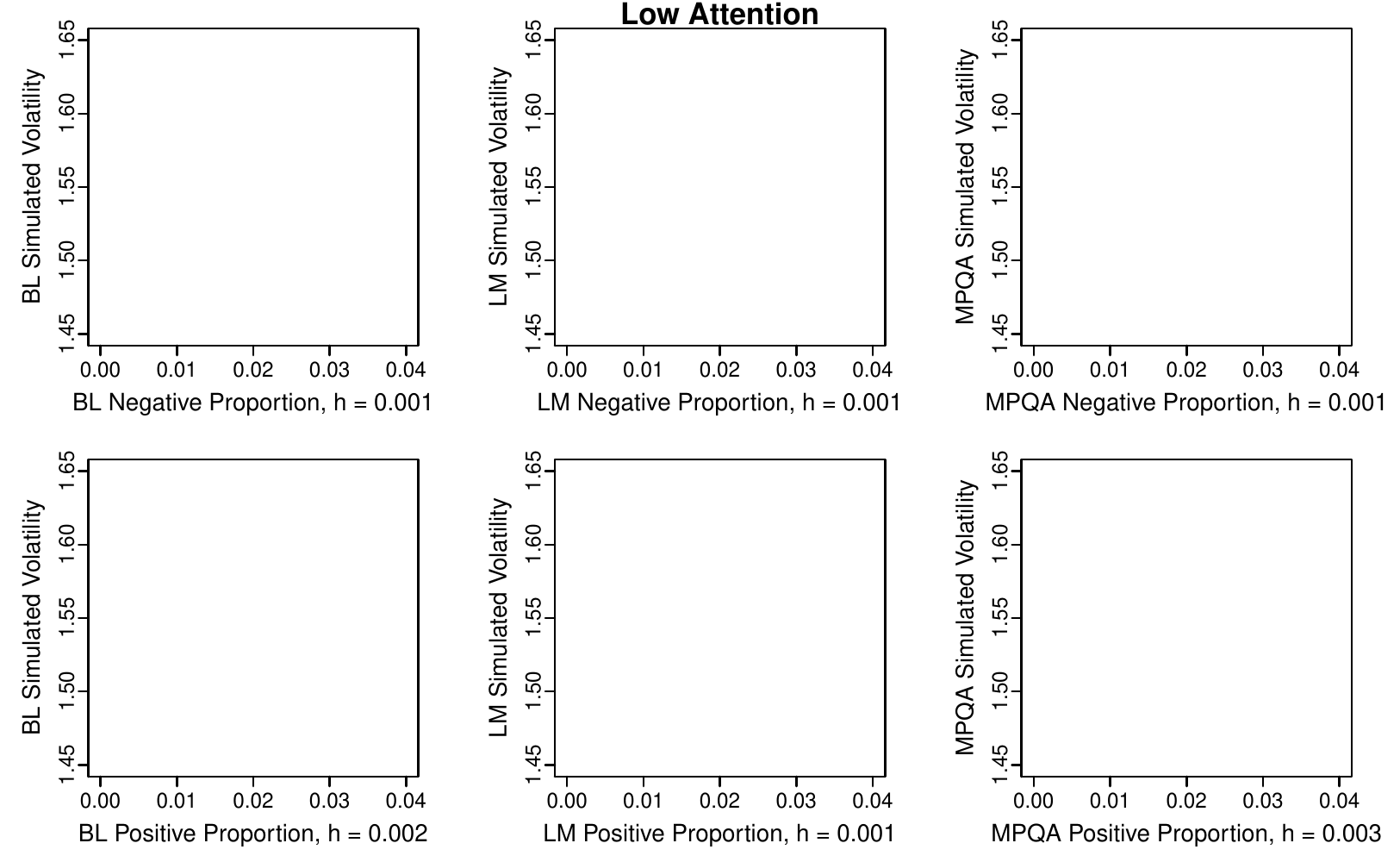}};
\end{tikzpicture}
\vspace{-6pt}
\caption{Monte Carlo Simulation based on Attention Analysis Results}
\label{fig:fig_attention_scat_selection}
\end{figure}

Like Section \ref{sect:MonteCarloSim}, we present a realistic Monte Carlo scenario for different attention groups using the results from Table \ref{tab:AttentionAnalysis}. We keep the parameter settings of the data generation and the calculation of confidence bands as before. Figure \ref{fig:fig_attention_scat_selection} summarizes the associations between the negative proportions and the simulated future volatilities across different attention groups. The scatter plots of the high attention panel are quite dense, whereas those of the low attention group are sparser due to its lower frequency of articles. Interestingly, the higher volatilities of high attention firms are prominently driven by negative text sentiment, but have an inverse relationship with positive sentiment. Through comparison of the confidence bands we can conclude for all three lexica that the effect of negative sentiment significantly differs from that of positive sentiment. The regions where the bands do not overlap are quite large for BL ($0.022$ - $0.056$) and MPQA ($0.020$ - $0.053$) but much smaller for LM ($0.019$ - $0.024$). The associations in the low attention panel are somewhat ambiguous. Indeed, we can note that the confidence bands for positive and negative sentiment overlap over the whole range of sentiment value and across all three lexica.  These simulations support the estimations in Table \ref{tab:AttentionAnalysis} with a strong link found in the ``extremely high" and ``high" attention groups and a preeminent asymmetric response.  The firms that have been paid high attentions are more sensitive to the text sentiment than negligible firms. The sentiment effect together with the observable asymmetry are highly influential on stock returns, volatilities and trading volumes. In this sense, their stock reactions are more responsive to the opinions in social media.  In other words, they are also more vulnerable to signals from small investors.

\subsection{Sector Analysis}
\label{sect:sect}
 The stock reactions that we analyze in relation to text sentiment can be further segmented into sector specific responses. Given a growing body of literature that has suggested that industry plays a role in stock reactions (see \cite{Fama&French:1997}, \cite{Chen&etal:2007}, \cite{Hong&etal:2007}), we investigate whether this relation is industry-specific in nature.  A detailed analysis of sector specific reactions would go far beyond the scope of this paper and is in fact the subject of research by \if0\blind{\cite{Bommes&etal:2015}}\fi \if1\blind{BLIND ITEM}\fi. We therefore only highlight a few insights from lexical sentiment for the business sectors. We ignore the ``Telecommuication Services" sector since it only contains two stock symbols.  Descriptive statistics for the other 8 sectors are displayed in Table \ref{tab:sectorstat} across the three lexical projections. It is of interest to study the variation of the proportion of negative over positive sentiments across the 8 sectors. One observes that consistently over all lexical projections the financial sector has the highest average discrepancy in negative and positive proportion. By contrast the health care sector has (except for MPQA) the lowest average discrepancy. Investors show their discrepant opinions or disagreement in a very extreme case of $Neg>Pos=0.5$, implying that 50\% of investors stand on one side and the rest of 50\% stand on the opposite side. Table \ref{tab:sectorstat} indicates that the financial sector related texts are more divergent in opinions than others and that apparently the health care sector does not receive such adverse opinion positions as the other sectors do. The investors who invest the stocks in health care sector are more likely to reach their shared concensus or convergent agreement.

 \FloatBarrier
\begin{summarytable}{h}
	\scalebox{0.71}{
		\centering
		\begin{tabular}{l|SSSSSSSSSS}
\hline \hline
			& \multicolumn{3}{c}{BL}
			& \multicolumn{3}{c}{LM} & \multicolumn{3}{c}{MPQA}	& {Attention} \\
			{Sector} & {$\mu_{Pos}$} & {$\mu_{Neg}$} & {$Neg>Pos$} & {$\mu_{Pos}$} & {$\mu_{Neg}$} & {$Neg>Pos$} & {$\mu_{Pos}$} & {$\mu_{Neg}$} & {$Neg>Pos$} & {Ratio}\\ \hline
			Consumer Discretionary  & 0.034 & 0.014 & 0.088 & 0.014 & 0.011 & 0.346 & 0.038 & 0.012 & 0.030 & 0.332 \\
			Consumer Staples  & 0.034 & 0.014 & 0.099 & 0.014 & 0.012 & 0.365 & 0.037 & 0.013 & 0.025 & 0.324 \\
			Energy  & 0.028 & 0.015 & 0.152 & 0.011 & 0.011 & 0.467 & 0.038 & 0.014 & 0.033 & 0.370 \\
			Financials  & 0.032 & 0.019 & 0.195 & 0.013 & 0.018 & 0.594 & 0.038 & 0.015 & 0.045 & 0.413 \\
			Health Care  & 0.035 & 0.014 & 0.059 & 0.014 & 0.011 & 0.344 & 0.039 & 0.014 & 0.031 & 0.287 \\
			Industrials  & 0.035 & 0.012 & 0.069 & 0.013 & 0.011 & 0.355 & 0.041 & 0.011 & 0.018 & 0.336 \\
			Information Technology  & 0.033 & 0.015 & 0.101 & 0.014 & 0.012 & 0.373 & 0.038 & 0.023 & 0.012 & 0.364 \\
			Materials  & 0.034 & 0.014 & 0.097 & 0.013 & 0.013 & 0.498 & 0.039 & 0.031 & 0.013 & 0.287 \\ 		
			\hline
			\hline
\multicolumn{11}{l}{\parbox{8.5in}{\vspace{3pt}Note: This table reports, for the BL, LM and MPQA methods, the mean values of positive ($\mu_{Pos}$) and ($\mu_{Neg}$) negative sentiment proportions as well as the proportion of relevant symbol-day combinations with dominance of negative sentiment. For each sector, an article is accumulated only if a firm appeared in this article belongs to this sector. The attention ratio for each sector is calculated as the number of days with articles related to this sector divided by the total number of days in the sample period.}} \\
		\end{tabular}	
	}
	\vspace{-6pt}
	\caption{Summary statistics in each sector}
\label{tab:sectorstat}
\end{summarytable}
\FloatBarrier

The attention also vary with the sectors. The evidence that financials sector has attracted the highest attention with an attention ratio of 0.413 may be attributed to (1) the investors' widespread involvement in this industry because we all need to keep a relationship with banks to deposit our money, trade for securities or some financial reasons; (2) the outbreak of the US subprime crisis and the European sovereign debt crisis have brought the highest attention to this sector; (3) their sensitivity on changes in the economy, monetary policy and regulatory policy. The health care sector, however, is much less attractive and this could be explained by a stable demand and reduced sensitivity to economic cycles. Given these observations we will now continue our analysis of stock reactions for these two sectors only, and leave a bundle of interesting issues to further research.

To address the important question of whether there is a sector dependent stock reactions, we further analyze how the text sentiment affects, as reported in Table \ref{tab:SectorAnalysis}, the future volatility, trading volume and return. In order to do so we employ the panel regression (as described in (\ref{eq:PanelVolatility})-(\ref{eq:PanelReturns})) and report the results in Table \ref{tab:SectorAnalysis}. The variable $I_{i,t}$ was used to indicate arrival of articles on this sector. Contingent on arriving articles, the three sentiment projections in financial sectors yielded significant and positive effects on future log volatility from negative proportions, meaning that increasing the negative text sentiments will result in higher volatility. The exclusive response to negative sentiment in financial sector indeed is in line with our entire panel evidence.
However, the finding in the health care sector is too insignificant to claim it. Potentially, investor inattention for the health care sector may cause a significant mispricing on the stocks. Investors possibly neglect the news of this sector posted on social media, or this sector has a slow information diffusion that could lead to a delayed reaction.
			
\FloatBarrier
\begin{regtable5}{ht}
                \scalebox{0.65}{
                               \centering
                               \begin{tabular}{l|SSSSSSSSS}
                                               \hline \hline
                                               & \multicolumn{3}{c}{BL}& \multicolumn{3}{c}{LM}& \multicolumn{3}{c}{MPQA} \\
                                               {Sector} & {$I_{i,t}$} & {$Pos_{i,t}$} & {$Neg_{i,t}$} & {$I_{i,t}$} & {$Pos_{i,t}$} & {$Neg_{i,t}$} & {$I_{i,t}$} & {$Pos_{i,t}$} & {$Neg_{i,t}$} \\ \hline
			& \multicolumn{9}{c}{Panel A: Future Volatility $\log \sigma_{i,t+1}$} \\
  Financials & -0.023 & -0.052 & 1.075$^{**}$ & -0.025 & 0.275 & 1.027$^{***}$ & -0.025 & -0.143 & 1.816$^{***}$ \\
  & (0.026) & (0.319) & (0.435) & (0.027) & (0.924) & (0.259) & (0.029) & (0.503) & (0.586) \\
  Health Care & 0.031 & -0.426 & -0.509 & 0.009 & 0.052 & -0.130 & 0.001 & -0.118 & 0.854 \\
  & (0.026) & (0.522) & (0.891) & (0.023) & (1.138) & (0.921) & (0.024) & (0.595) & (0.783) \\\hline
			& \multicolumn{9}{c}{Panel B: Future Detrended Log Trading Volume $V_{i,t+1}$} \\
  Financials & 0.037$^{*}$ & -0.334 & 0.015 & 0.017 & 1.110 & -0.313 & 0.054$^{***}$ & -0.747$^{**}$ & 0.049 \\
  & (0.020) & (0.494) & (0.527) & (0.015) & (0.766) & (0.476) & (0.015) & (0.305) & (0.536) \\
  Health Care & 0.031 & 0.110 & -0.314 & 0.022 & 0.603 & -0.042 & 0.037 & -0.104 & -0.211 \\
  & (0.023) & (0.436) & (0.846) & (0.018) & (0.863) & (0.837) & (0.025) & (0.443) & (0.873) \\ \hline
			& \multicolumn{9}{c}{Panel C: Future Returns $R_{i,t+1}$} \\
  Financials & -0.001 & 0.034$^{*}$ & 0.028$^{**}$ & -0.000 & 0.030 & 0.042$^{**}$ & 0.001 & 0.003 & 0.013 \\
  & (0.001) & (0.017) & (0.014) & (0.001) & (0.033) & (0.016) & (0.001) & (0.020) & (0.019) \\
  Health Care & 0.000 & -0.000 & 0.008 & 0.000 & 0.006 & 0.015 & 0.000 & 0.006 & -0.011 \\
  & (0.000) & (0.008) & (0.018) & (0.000) & (0.019) & (0.018) & (0.001) & (0.012) & (0.022) \\
			\hline \hline
\multicolumn{10}{l}{\parbox{9.5in}{\vspace{4pt}$^{***}$ refers to a $p$ value less than 0.01, $^{**}$ refers to a $p$ value more than or equal to 0.01 and smaller than 0.05, and $^{*}$ refers to a $p$ value more than or equal to 0.05 and less than 0.1. Values in parentheses are standard errors.}} \\
		\end{tabular}		
	}
	\vspace{-6pt}
	\caption{Sector analysis: The Impact on future Volatility, Trading Volume and Returns}
\label{tab:SectorAnalysis}
\end{regtable5}
\FloatBarrier

The trading volume is another stock reaction we may attribute to text sentiments. Using the BL and the MPQA projection method, we find that the arrival of article brings relevant information and therefore stimulates the trading volume. It is interesting to note that contingent on arriving articles, the negative sentiment distilled using the BL and LM methods is significantly positively related to stock returns on the next trading day. To investigate the reason for this, we also run a contemporaneous regression for the financials sector (results not shown) and found a significantly negative impact of the negative sentiment distilled using the BL and MPQA methods on contemporaneous returns $R_{i,t}$, and the size of the coefficients is about twice of that in lagged regression in Table \ref{tab:SectorAnalysis}. This might suggest that the market participants monitor financial companies quite carefully and overreact in case of bad news. On the next day, the participants fully recognize the scope of the news and reverse part of their prior decisions, and hence the negative sentiment on trading day $t$ has positive impact on returns on trading day $t+1$. This is also in line with the finding in \cite{Kuhnen:2015} which suggests that that being in a negative domain leads people to form overly pessimistic beliefs about stocks. After the 2008 financial crisis and the bankruptcy of some major financial companies, this might be the case for the financials sector.

From these analysis, we know that investors indeed pay different attentions to sectors they are of interest, and their attentions effectively govern the equity's variation. Attention constraints in some sectors may affect investors' trading decisions and the speed of price adjustments.

\section{Conclusion}
\label{sect:Conclusion}

In this paper, to analyze the reaction of stocks' future log volatility, future detrended log trading volume and future returns to social media news, we distill sentiment measures from news using two general-purpose lexica (BL and MPQA) and a lexicon specifically designed for financial applications (LM). We demonstrate that these sentiment measures carry incremental information for future stock reactions. Such information varies across lexical projections, across groups of stocks that attract different level of attention, and across different sectors. The positive and negative sentiments also have asymmetric impact on future stock reaction indicators. A detailed summary of the results is given in Table \ref{tab:ResultsSummary} in the Supplementary Material. There is no definite picture for which lexicon is the best. This is an important contribution of our paper to the line of research on textual analysis for financial market. Besides, the advanced statistical tools that we have utilized, including panel regression and confidence bands, are novel contributions to this line of research.


\bibliographystyle{apalike}
\bibliography{Zha_Hae_Che_Bom_Distillation_of_News_Flow_into_Analysis_of_Stock_Indicators}

\section{Supplementary Material}

Table \ref{tab:ResultsSummary} summarizes all the results from entire panel sample analysis, attention analysis and sector analysis. Take the ``BL" row in Panel A as an example. Arrival of articles ($I_{i,t}$) and the positive sentiment distilled using the BL method ($Pos_{i,t}$) has no significant impact on future volatility $\log \sigma_{i,t+1}$ in entire sample analysis, attention analysis or sector analysis; the negative sentiment distilled using the BL method ($Neg_{i,t}$) is significantly positively related to future volatility in entire sample analysis and for the ``Extremely High" group in attention analysis, and is significantly negatively related to future volatility for the ``Health Care" sector in sector analysis.
\begin{regtable5}{ht}
		\centering
\scalebox{0.55}{
		\begin{tabular}{l|l|c|c|c} \hline \hline
Lexicon & Type of Analysis & $I_{i,t}$ & $Pos_{i,t}$ & $Neg_{i,t}$ \\\hline
			\multicolumn{5}{c}{Panel A: Future Volatility $\sigma_{i,t+1}$} \\\hline
\multirow{3}*{BL}& Entire Sample & / & Negative & Positive\\
& Attention Analysis & / & / & Positive for ``Median", ``High" and ``Extremely High"\\
& Sector Analysis & / & / & Positive for ``Financials"\\\hline
\multirow{3}*{LM}& Entire Sample & Negative & / & Positive\\
& Attention Analysis & Negative for ``High" & / & Positive for ``Median", ``High" and ``Extremely High"\\
& Sector Analysis & / & / & Positive for ``Financials" \\\hline
\multirow{3}*{MPQA}& Entire Sample & / & Negative & Positive\\
& Attention Analysis & / & Negative for ``Median" and ``High" & Positive for ``Median", ``High" and ``Extremely High"\\
& Sector Analysis & / & / & Positive for ``Financials" \\\hline
		\multicolumn{5}{c}{Panel B: Future Detrended Log Trading Volume $V_{i,t+1}$} \\\hline
\multirow{3}*{BL}& Entire Sample & Positive & Negative & Positive\\
& Attention Analysis & Positive for ``low", ``Median" and ``High" & Negative for ``Median" & Positive for ``Median" and ``Extremely High"\\
& Sector Analysis & Positive for ``Financials" & / & /\\\hline
\multirow{3}*{LM}& Entire Sample & Positive & / & Positive\\
& Attention Analysis & Positive for all groups& Positive for ``High" & Positive for ``Extremely High"\\
& Sector Analysis & / & / & /\\\hline
\multirow{3}*{MPQA}& Entire Sample & Positive & Negative & Positive\\
& Attention Analysis & Positive for all groups & Negative for ``Median" & Positive for ``Extremely High"\\
& Sector Analysis & Positive for ``Financials" & Negative for ``Financials" & /\\\hline
			\multicolumn{5}{c}{Panel C: Future Returns $R_{i,t+1}$} \\\hline
\multirow{3}*{BL}& Entire Sample & / & Positive & /\\
& Attention Analysis & / & Positive for ``Median" and ``High" & Negative for ``High"\\
& Sector Analysis & /& Positive for ``Financials"&Positive for ``Financials"\\\hline
\multirow{3}*{LM}& Entire Sample & / & Positive  & /\\
& Attention Analysis & Positive for ``High" & Positive for ``Median" and ``High" & Negative for ``High", positive for ``Extremely High"\\
& Sector Analysis & /& /&Positive for ``Financials"\\\hline
\multirow{3}*{MPQA}& Entire Sample & / & Positive & /\\
& Attention Analysis & Positive for ``Extremely High" & Positive for ``Median" and ``High" & Negative for ``High"\\
& Sector Analysis & /& /&/\\\hline\hline
\multicolumn{5}{l}{\parbox{9.5in}{\vspace{4pt}The signs of the significant coefficients are given, with a significance level of 0.1.}} \\
		\end{tabular}	
}
	\vspace{-6pt}
	\caption{Summary of the Results}
	\label{tab:ResultsSummary}
\end{regtable5}

\end{document}